\documentclass[12pt]{article}
\usepackage[colorlinks=true,linkcolor=blue,urlcolor=blue,filecolor=black,citecolor=red,pdfstartview=FitV,pdftitle={},pdfsubject={},
pdfkeywords={},pdfpagemode=None,bookmarksopen=true]{hyperref}
\usepackage{graphicx}
\usepackage{epstopdf}%
\usepackage{amsmath}
\usepackage{amsfonts}
\usepackage{amssymb}
\usepackage{color}%
\usepackage{dcolumn}
\usepackage{slashed}
\usepackage{amssymb,ulem}
\usepackage{float}
\usepackage{amsthm,amsmath,amssymb}
\usepackage{mathrsfs}
\usepackage{subfigure}
\usepackage{wrapfig}
\usepackage{indentfirst}
\providecommand{\U}[1]{\protect\rule{.1in}{.1in}}

\renewcommand{\thefootnote}{\fnsymbol{footnote}}

\begin{document}

\vspace{12mm}

\begin{center}
{{{\Large {\bf Nonlinearly scalarized rotating black holes in Einstein-scalar-Gauss-Bonnet theory }}}}\\[10mm]

{Meng-Yun Lai$^{a}$\footnote{mengyunlai@jxnu.edu.cn;}, De-Cheng Zou$^{a,b}$\footnote{Corresponding author. dczou@jxnu.edu.cn;}, Rui-Hong Yue$^b$\footnote{rhyue@yzu.edu.cn;}, \\ and
Yun Soo Myung$^c$\footnote{ysmyung@inje.ac.kr;}}\\[8mm]

{${}^a$College of Physics and Communication Electronics, Jiangxi Normal University, \\ Nanchang 330022, China\\[0pt] }
{${}^b$Center for Gravitation and Cosmology and School of Physical Science and Technology, Yangzhou University, Yangzhou 225009, China\\[0pt]}
{${}^c$Institute of Basic Sciences and Department  of Computer Simulation,\\ Inje University, Gimhae 50834, Korea\\[0pt] }
\end{center}

\vspace{2mm}

\begin{abstract}
In this paper, we discuss a fully nonlinear mechanism for the formation of scalarized rotating black holes in Einstein-scalar-Gauss-Bonnet gravity, where Kerr black holes are linearly stable, but unstable against nonlinear scalar perturbations. With the help of the pseudospectral method, we obtain a spectrum of nonlinearly scalarized rotating black hole solutions with multiple scalarized branches. Moreover, we investigate the thermodynamic properties of nonlinearly scalarized rotating black holes and find the phase transition between Kerr and these scalarized black holes.
\end{abstract}
\vspace{5mm}

\newpage
\renewcommand{\thefootnote}{\arabic{footnote}}
\setcounter{footnote}{0}

\section{Introduction}

As a consequence of the ``no-hair theorems'', the general relativity black holes
are described three observables of mass $M$, electric charge $Q$ and rotation parameter $a=J/M$ \cite{Carter1971zc,Ruffini1971bza}. It rules out a black hole with conformal scalar hair in asymptotically flat spacetimes when accounting for the divergence of a scalar field on the horizon \cite{Bekenstein:1974sf}-\cite{Bronnikov:1978mx}. Therefore, the ``no-hair theorems'' gradually become a major obstacle to our physical hopes of discovering new fundamental fields that interact with the curved black hole spacetimes.

Nevertheless, one may circumvent ``no-hair theorems'' by violating some of their underlying assumptions. For instance,
Damour and Esposito-Farese~\cite{Damour:1993hw,Damour:1996ke} first recovered a mechanism of spontaneous scalarization in scalar-tensor theory when studying neutron stars. Recently, Doneva and Yazadjie \cite{Doneva:2017duq} constructed realistic scalarized neutron star solutions in Einstein-scalar-Gauss-Bonnet (EsGB) gravity with a nontrivial coupling of a scalar field $\phi$ to the Gauss--Bonnet(GB) curvature term $R^2_{\rm GB}$. Similar discussions have been also extended to the black holes. Doneva and Yazadjiev{et al}.\cite{Doneva:2017bvd} found that the spontaneous scalarization may take place around Schwarzaschild balck holes,
due to a tachyonic instability triggered by the coupling of scalar field to the GB term which is analogous to that inside relativistic
stars triggered by the coupling of scalar field to the matter field. Below certain mass, the Schwarzschild solution becomes
unstable and new branch of solutions with nontrivial scalar field bifurcate from the Schwarzschild one. In particular,
Antoniou \textit{et al}. \cite{Antoniou:2017acq} asserted that a regular scalarized black holes can arise as the result of the
synergy between only the scalar field function $f(\phi)$ and the GB term in the EsGB theory by evaluating the asymptotic forms of the energy-momentum tensor near the horizon and at infinity. Then, existing ``no-hair theorems'' are easily evaded. Until now, the phenomenon of black hole spontaneous scalarization has received a lot of attention in the EsGB gravity \cite{Silva:2017uqg}-\cite{Myung:2018iyq}. These theories possess black holes with scalar hair, whose
properties have been investigated in great details~\cite{Minamitsuji:2018xde}-\cite{Blazquez-Salcedo:2021npn}.
In addition, Ref.~\cite{Blazquez-Salcedo:2018jnn} pointed out that, under radial perturbations, the scalarized black holes are unstable for a quadratic coupling, whereas it is stable for an exponential form in the EsGB theory.  Motivated by current and future
gravitational wave observations from black hole mergers, the axial~\cite{Blazquez-Salcedo:2020rhf}
and polar~\cite{Blazquez-Salcedo:2020caw} perturbations of scalarized black holes have
been investigated  to obtain the quasinormal modes (QNMs) in the EsGB theory since
QNMs could describe the ringdown after merging.

Recently, the phenomenon of spontaneous scalarization
of spinning black holes has been attractive to the readers.
Dima \textit{et al}.~\cite{Dima:2020yac} first discovered that the high rotation  can induce
tachyonic instability of Kerr black hole for a positive coupling  by evaluating the $(1+1)$-dimensional scalar evolution equation in
EsGB theory.  When choosing a negative coupling, an upper $a$-bound ($a/M\ge 0.5$) comes out as the onset of scalarization for Kerr black hole,
but the low rotation ($a/M<0.5$) is supposed to  suppress spontaneous scalarization.
Shortly afterward, the critical rotation parameter $(a/M)_c=0.5$ for Kerr black holes was  computed analytically~\cite{Hod:2020jjy}
and numerically~\cite{Zhang:2020pko}-\cite{Berti:2020kgk} in the EsGB theory with negative couplings.
In this direction, spin-induced scalarized black holes have been also constructed numerically in the EsGB theory with positive
coupling~\cite{Cunha:2019dwb}-\cite{Herdeiro:2020wei}. Zou and Myung \cite{Zou:2021ybk} have also discussed spontaneous
scalarization of Kerr black holes by including two different coupling functions. These imply that the rotation parameter $a$ and the
coupling parameter $\alpha$ are key factors for achieving spontaneous scalarization of spinning black holes.

It is interesting to point out that the most studied driving mechanism leading to
spontaneous scalarization in the  EsGB theory is a tachyonic instability due to an effective negative
mass squared ($\mu_{eff}^2\sim F''(\phi)R^2_{GB}$) for the scalar
field. Recently, Doneva and Yazadjiev \cite{Doneva:2021tvn} found when the coupling function
is chosen such that the effective mass is zero, such as coupling functions takes the forms
\begin{eqnarray}
F_1(\phi)=\frac{1}{4\kappa}(1-e^{-\kappa\phi^4}),\quad F_2(\phi)=\frac{1}{6\kappa}(1-e^{-\kappa\phi^6}),
\end{eqnarray}
Schwarzschild black holes always hold stable under linear scalar field perturbation. However,
the Schwarzschild black holes become unstable against nonlinear scalar perturbations if the amplitude of the perturbations is large enough.
Moreover, the obtained in this way scalarized phases are not continuously connected to the Schwarzschild black hole. Depending on the parameter $\kappa$ in the coupling functions, three branches of scalarized phases can exist and the stable scalarized phase has the largest entropy among all the branches of hairy black holes. Also, one of two nonlinearly scalarized black holes is stable against the radial perturbations~\cite{Blazquez-Salcedo:2022omw}.
Later, Doneva \textit{et al}.~\cite{Doneva:2022yqu} further discovered that in the EsGB theory, the Kerr black hole is stable under linear perturbations, but it is unstable  against
larger nonlinear perturbations. By evolving in time the nonlinear scalar field equation on the Kerr background, it turns out that there is a threshold amplitude of the scalar perturbation above which the Kerr black hole loses the linear stability and scalarized rotating black holes could form. The scalarization of charged black holes has been also
studied in the Einstein-Maxwell-scalar gravity coupled with a nontrivial coupling of a
scalar field function $f(\phi)=1+\alpha\phi^4$ to the Maxwell term $F_{\mu\nu}F^{\mu\nu}$\cite{Blazquez-Salcedo:2020nhs}.
Inspired by these works, we will further derive the solutions for nonlinearly scalarized rotating black holes in the EsGB theory.
Moreover, the full nonlinear and self-consistent analysis of these black
holes will show the existence of a spectrum of solutions with  multiple scalarized black hole branches.
Importantly, we will investigate the thermodynamic property for nonlinearly scalarized rotating black holes to  explore  a phase transition between Kerr and these black holes.

The plan of our work is as follows. In Sec.~\ref{sec2}, we  mention briefly the nonlinearized scalar perturbation on the
Kerr black hole in the EsGB theory. By making use of the pseudospectral method, we will construct numerical solutions
of nonlinearly scalarized rotating  black holes in Sec.~\ref{sec3}.
Sec.~\ref{sec4} is devoted to investigating physical and thermodynamic  properties of these black holes.
Finally, we close the paper with a discussion and conclusions in Sec.~\ref{sec5}.

\section{NONLIEAR INSTABLITY OF KERR BLACK HOLES}  \label{sec2}

The action of Einstein-scalar-Gauss-Bonnet gravity reads as
\begin{eqnarray}
{\cal S}_{\rm EsGB}\equiv\int d^4 x\sqrt{-g}{\cal L} =\frac{1}{16 \pi}\int d^4 x\sqrt{-g}\left( R-2\partial_\mu \phi \partial^\mu \phi
+\lambda^2 F(\phi) R^2_{\rm GB}\right),\label{Action1}
\end{eqnarray}
where  $F(\phi)$ is  the coupling function
and $\lambda$ is a scalar coupling parameter to the Gauss--Bonnet term as
\begin{eqnarray}
{\cal R}^2_{\rm GB}=R^2-4R_{\mu\nu}R^{\mu\nu}+R_{\mu\nu\rho\sigma}R^{\mu\nu\rho\sigma}.\label{GBterm}
\end{eqnarray}

Varying the action (\ref{Action1}) with scalar $\phi$ and metric $g_{\mu\nu}$, one obtains two field equations
\begin{eqnarray}
&&\square \phi +\frac{\lambda^2}{4}F'(\phi) {\cal R}^2_{\rm GB}=0, \label{s-equa}\\
&&E_{\mu\nu}=R_{\mu\nu}-\frac{1}{2}R g_{\mu\nu}+\Gamma_{\mu\nu}-T^{\phi}_{\mu\nu}=0, \label{equa1}
\end{eqnarray}
where
\begin{eqnarray}
\Gamma_{\mu\nu}&\equiv&-2R\nabla_{(\mu}\phi_{\nu)}-4\nabla_{\sigma}\phi^{\sigma}\left(R_{\mu\nu}
-\frac{1}{2}R g_{\mu\nu}\right)+4R_{\mu\sigma}\nabla^\sigma\phi_\nu\nonumber\\
&&+4R_{\nu\sigma}\nabla^\sigma\phi_\mu-4g_{\mu\nu}R^{\sigma\rho}\nabla_\sigma\phi_\rho
+4R^\sigma_{~\mu\rho\nu}\nabla^\rho\phi_\sigma, \label{gamma}\\
T^{\phi}_{\mu\nu}&=&2\nabla_\mu \phi\nabla_\nu \phi -(\nabla\phi)^2g_{\mu\nu}\label{Tphi}
\end{eqnarray}
with $\phi_\mu\equiv\lambda^2 F'(\phi)\nabla_\mu\phi$.

Notice that the EsGB gravity in Eq.~(\ref{Action1}) admits the Kerr black hole solution with a  vanishing scalar
($\phi=0$) and the coupling function $F(0)=0$.
Early works \cite{Dima:2020yac}-\cite{Berti:2020kgk} pointed out that $\mu_{eff}^2=-\frac{\lambda^2}{4}F''(0)R^2_{\rm GB}$
can be regarded as an effective mass squared of scalar perturbation on a fixed background (Kerr black hole).
It might trigger a tachyonic (linear) instability when either $F''(0)<0$
or $F''(0)>0$.
This process is named the spontaneous scalarization for a Kerr black hole. However, if the coupling function $F(\phi)$ takes the form
\begin{eqnarray}
F(\phi)=\frac{1}{4\kappa}\left(1-e^{-\kappa\phi^4}\right),\label{funterm}
\end{eqnarray}
we have the properties
\begin{eqnarray}
F(0)=0,\quad F'(0)=0, \quad F''(0)=0.
\end{eqnarray}
Here $\kappa$ is regarded as  a coupling parameter.
 The linearized scalar equation around the Kerr black hole background leads to
\begin{eqnarray}
\bar{\square}_{\rm K} \delta \phi=0
\end{eqnarray}
which is a massless scalar equation.
It implies that there is no tachyonic instability anymore for Kerr black hole because of $\mu_{eff}^2=0$. Interestingly, Ref.~\cite{Doneva:2022yqu} has  pointed out that the Kerr black hole becomes unstable against nonlinear scalar perturbations
if a large  initial perturbation is imposed and  the nonlinear instability can lead to the formation of nonlinearly scalarized rotating black holes. In other words, a newly nonlinear scalarization occurs, being distinct from spontaneous scalarization of the Kerr black hole~\cite{Dima:2020yac}-\cite{Berti:2020kgk}. In the following sections,
we turn to solve a fully nonlinear coupled system of field equations by using the pseudospectral method and obtain a clear picture of
fully nonlinear scalarization for Kerr black holes in the EsGB gravity. Note that we choose the coupling parameter $\lambda=1$ in the numerical results presented below.

\section{NONLINEARLY SCALARIZED ROTATING BLACK HOLES} \label{sec3}

First of all, we introduce the stationary and axisymmetric metric ansatz written in quasi-isotropic coordinates \cite{Fernandes:2022gde}
\begin{eqnarray}
ds^2_{\rm QI}=-f N^2 dt^2+\frac{g}{f}\Bigg[h\left(dr^2+r^2d\theta^2\right)
+r^2\sin^2\theta\left(d\varphi-\frac{W}{r}(1-N)dt\right)^2\Bigg],\label{metric}
\end{eqnarray}
where the three spatial coordinates range over the intervals
\begin{eqnarray}
r\in[r_H,\infty ], \quad \theta\in[0,\pi], \quad \varphi\in[0,2\pi].
\end{eqnarray}
The full configuration of the black hole is therefore described by the functions of $(r, \theta)$: $f$, $g$, $h$ and $W$.

Assuming the line element \eqref{metric} to be a solution to the theory of gravity at hand, the
functions, $f$, $g$, $h$, $W$ and scalar field $\phi$ should satisfy a set of coupled partial differential equations (PDEs)
when substituting the metric ansatz \eqref{metric} into  Eqs.~(\ref{s-equa}) and (\ref{equa1}).
In the next sections, we will solve the four combinations of the Einstein equation which diagonalize the Einstein tensor with respect to the operator ($\partial_r^2+r^{-2}\partial^2_{\theta}$),
\begin{eqnarray}
&&E_{~\mu}^{\mu}-2E_{~t}^{t}-\frac{2W r_H}{r^2}E_{~t}^{\varphi}=0,\nonumber\\
&&E_{~t}^{\varphi}=0,\nonumber\\
&&E_{~r}^{r}+E_{~\theta}^{\theta}=0,\nonumber\\
&&E_{~\varphi}^{\varphi}-\frac{2W r_H}{r^2}E_{~t}^{\varphi}-E_{~r}^{r}-E_{~\theta}^{\theta}=0\label{eom}
\end{eqnarray}
and Klein-Gordon equation
\begin{eqnarray}
\square \phi +\frac{\lambda^2}{4}F'(\phi) {\cal R}^2_{\rm GB}=0.\label{KG}
\end{eqnarray}

In order to perform the numerical integration of Eqs.\eqref{eom}\eqref{KG}, we introduce a new radial coordinate for convenience,
\begin{eqnarray}
x\equiv1-\frac{2r_H}{r},\label{rx}
\end{eqnarray}
which maps $r\in[r_H,\infty]$ to $x\in[-1,1]$. Moreover, the suitable boundary conditions
should be imposed. At the event horizon $x=-1$, we adopt the boundary conditions with
\begin{eqnarray}
f-2\partial_x f=0, \quad g+2\partial_x g=0, \quad \partial_x h=\partial_x \phi=0,\quad W-\partial_xW=0. \label{EHBC}
\end{eqnarray}
 For asymptotically flat solutions,  one requires
\begin{eqnarray}
f=g=h=1, \quad \phi=0, \quad \partial_xW+j(1+\partial_xf)^2=0 \label{AsBC1}
\end{eqnarray}
at $x=1$, where the dimensionless spin $j$ is given by $j\equiv J/M^2=a/M$. For the axis boundary conditions, the axial symmetry and regularity impose the following boundary conditions on the symmetry axis
\begin{eqnarray}
\partial_\theta f=\partial_\theta g=\partial_\theta h=\partial_\theta W=\partial_\theta\phi=0, \quad {\rm for}  \quad \theta=0, \quad \frac{\pi}{2}, \label{AxBC1}
\end{eqnarray}
 because all solutions found in this work are symmetric with respect to a
reflection on the equatorial plane at $\theta=\pi/2$.  Therefore, it is sufficient to confine the range $\theta\in[0,\pi/2]$.
Additionally, the absence of conical singularities implies
\begin{eqnarray}
h=1, \quad {\rm for}  \quad \theta=0, \quad  \frac{\pi}{2}.
\end{eqnarray}

With the above boundary conditions, the system of coupled PDEs \eqref{eom} and \eqref{KG} can be numerically solved to obtain the nonlinearly scalarized rotating black hole solutions through the spectral method \cite{Fernandes:2022gde}. Now,  we briefly describe the steps. We first decompose the five functions to be solved (${\cal F}=\{f,g,h,W,\phi\}$) into radial and angular
parts, together with a suitable spectral expansion
\begin{eqnarray}
{\cal F}^{(k)} =\sum_{i=0}^{N_x-1}\sum_{j=0}^{N_\theta-1}\alpha_{ij}^{(k)}T_i(x)\cos(2j\theta),\label{spexpansion1}
\end{eqnarray}
where $T_i(x)$ is the Chebyshev polynomial, and $N_x$ and $N_\theta$ denote the resolutions in the radial and angular directions. Plugging the spectral expansions \eqref{spexpansion1} into the coupled field equations \eqref{eom} and \eqref{KG}, we can calculate the resulting equations at each Gauss-Chebyshev points  defined by
\begin{eqnarray}
x_k&=&\cos\Big[\frac{(2k+1)\pi}{2(N_x-2)}\Big], \quad k=0,..., N_x-3,\\
\theta_l&=&\frac{(2l+1)\pi}{4N_\theta}, \quad l=0,..., N_\theta-1.
\end{eqnarray}
Together with the boundary conditions, we end up with a nonlinear system of equations consisting of $N_{\cal F}\times N_x \times N_\theta$ equations with respect to the spectral coefficients $\{\alpha_{ij}^{(k)}\}$, where $N_{\cal F}=5$ is the number of the functions to be solved. Finally, the nonlinear system of equations  can be numerically solved through the Newton-Raphson method. The more details are presented in the Appendix.

\begin{figure*}[t!]
\centering
\subfigure[]{
\includegraphics[width=0.45\textwidth]{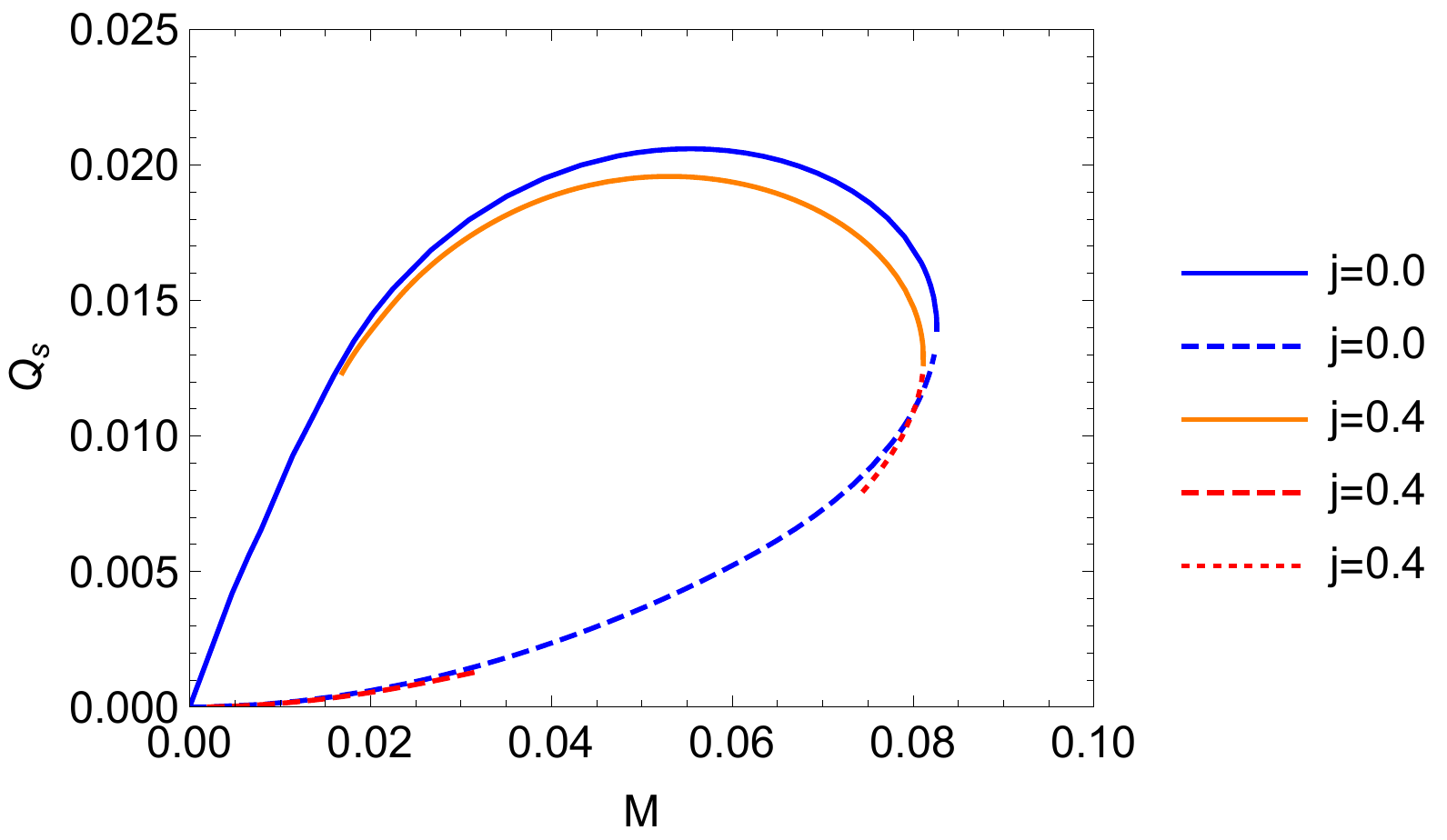}}
\hfill%
\subfigure[]{
\includegraphics[width=0.45\textwidth]{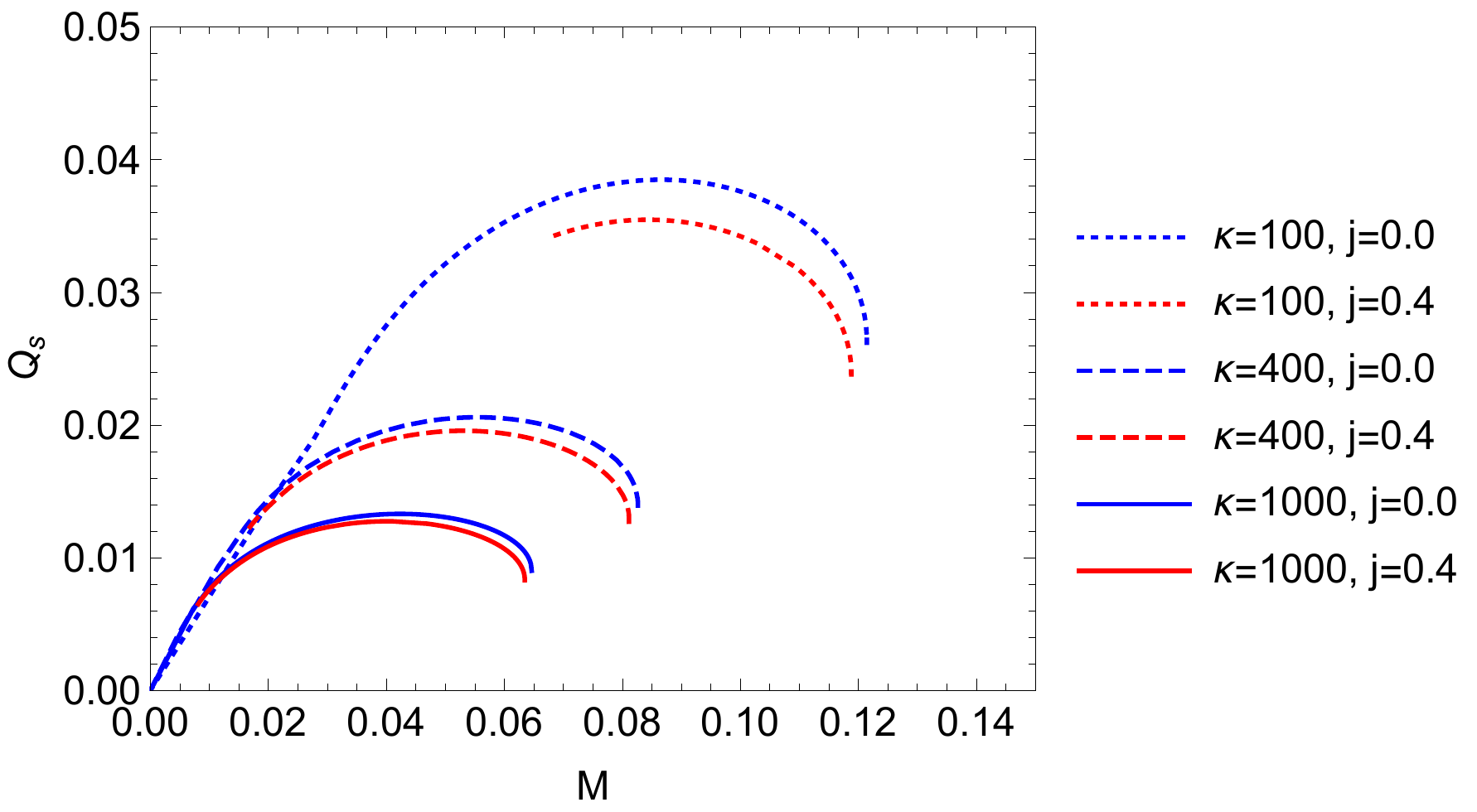}}
\caption{(a): Scalar charge $Q_s$ as function of black hole mass $M$ for static $(j=0)$ and scalarized rotating $(j=0.4)$ black holes with $\kappa=400$. (b): Scalar charge $Q_s$ as a function of black hole mass $M$ for other scalarized black holes with $\kappa=100, 400$ and 1000.  }\label{fig11a}
\end{figure*}

With these numerical solutions, the physical quantities of black holes such as mass $M$, angular momentum $J$, and scalar charge $Q_s$ can be expressed in terms of the coordinate $x$ as
\begin{eqnarray}
M&=&r_H(1+\partial_x f)|_{x=1}, \\
J&=&-r_H^2\partial_x W|_{x=1}, \\
Q_s&=& -2r_H\partial_x\phi|_{x=1}.\label{MJexp}
\end{eqnarray}
In Ref. \cite{Doneva:2021tvn}, Doneva and Yazadjiev discovered two branches for fully nonlinear scalarization of Schwarzschild black hole in the EsGB gravity. For scalarized rotating black holes, we find the existence of three branches, being different from the static case. The scalar charge $Q_s$ is plotted as a function of mass $M$ for the static and rotating scalarized black holes with parameter $\kappa=400$ in Fig.~\ref{fig11a}. As shown in Figure \ref{fig11a}(a), the branch 2 in static case is now divided into two sub-branches for rotating case, and the orange solid, red dashed and red dotted curves are called branch 1, branch 2a and branch 2b, respectively. Moreover, branch 1 and branch 2a are connected at finite mass $M=0.0811$.

In addition, there are three breakpoints in the curve of the rotating scalarized black holes compared to the static case. The numerical process no longer converges as it approaches the three breakpoints of the orange and red curves in Figure \ref{fig11a}(a), while we do not observe a singular behavior in the vicinities of these breakpoints. Actually, this phenomenon is quite common in EsGB models \cite{Herdeiro:2020wei}\cite{Kleihaus:2011tg}\cite{Sotiriou:2013qea}, and the black hole solutions corresponding to the breakpoints are called critical solutions in the literature. An explanation is based on the horizon expansions of the field equations \cite{Kleihaus:2015aje}\cite{Delgado:2020rev} with
\begin{eqnarray}
\phi(x,\theta)=\phi_0(\theta)+\phi_2(\theta)(x+1)^2+\cdots.
\end{eqnarray}
One can find a quadratic equation of $\phi_2$
\begin{eqnarray}
    \phi^2_2+p\phi_2+q=0,
\end{eqnarray}
where the coefficients $p$ and $q$  depend on the values of the metric functions and their derivatives at the horizon.
Therefore, a regular solution exists only if $\Delta=p^2-4q>0$. When the critical solution is exceeded, $\Delta$ becomes
negative and the regular hairy solution no longer exists.
In Figure \ref{fig11a}(b), we further show the rotating scalarized black holes with spin parameter $j=0.4$ and $\kappa=100$, 400 and 1000.
We find that rotating scalarized black holes with larger values of $\kappa$ possess smaller scalar charge $Q_s$ and mass $M$.
However, the curves of physical quantities for these scalarized black holes with different $\kappa$ are very similar.
Therefore, we will mainly consider the case of $\kappa = 400$ in the following sections.

\begin{figure*}[t!]
\centering
\subfigure[]{
\includegraphics[width=0.44\textwidth]{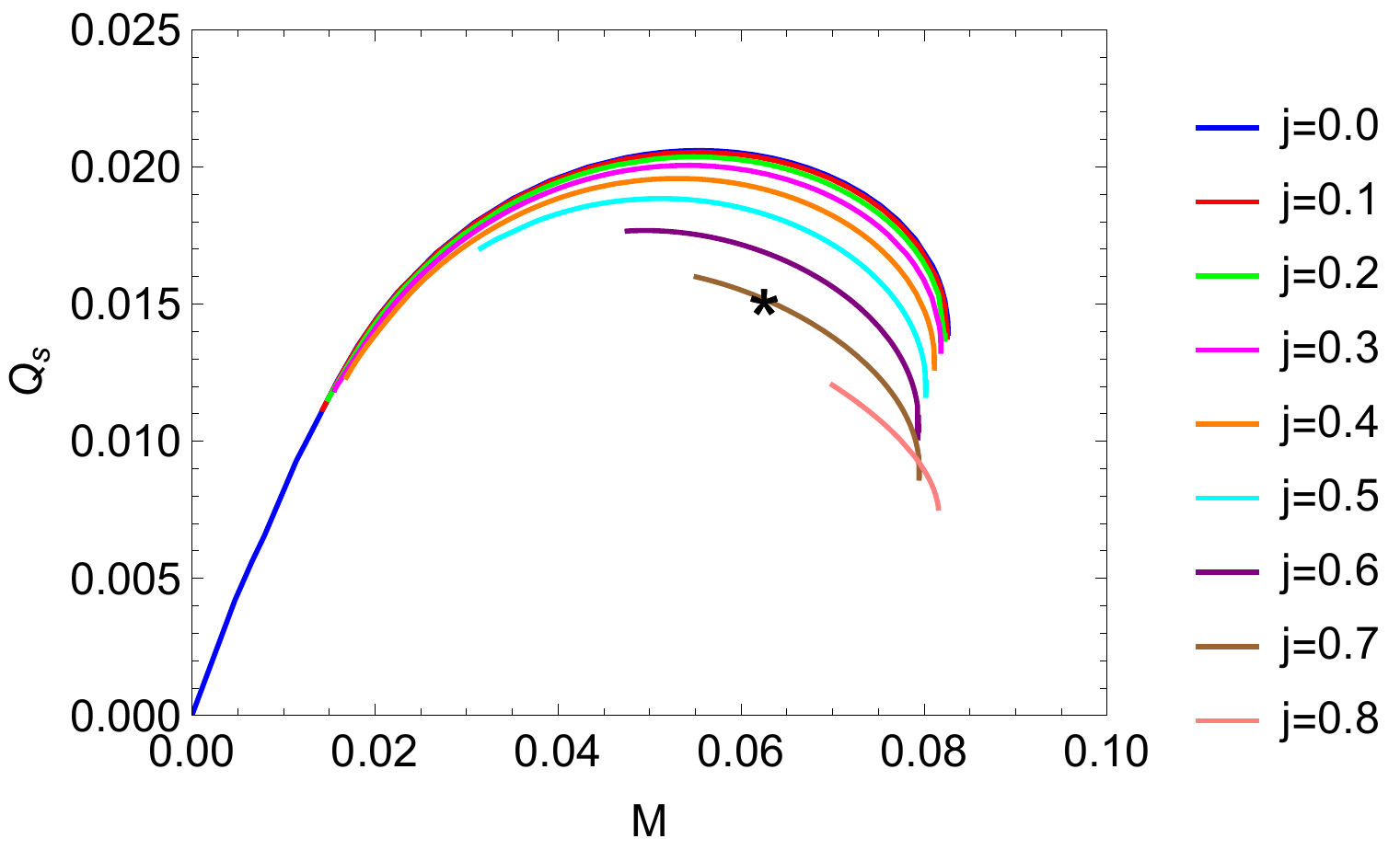}}
\hfill%
\subfigure[]{
\includegraphics[width=0.37\textwidth]{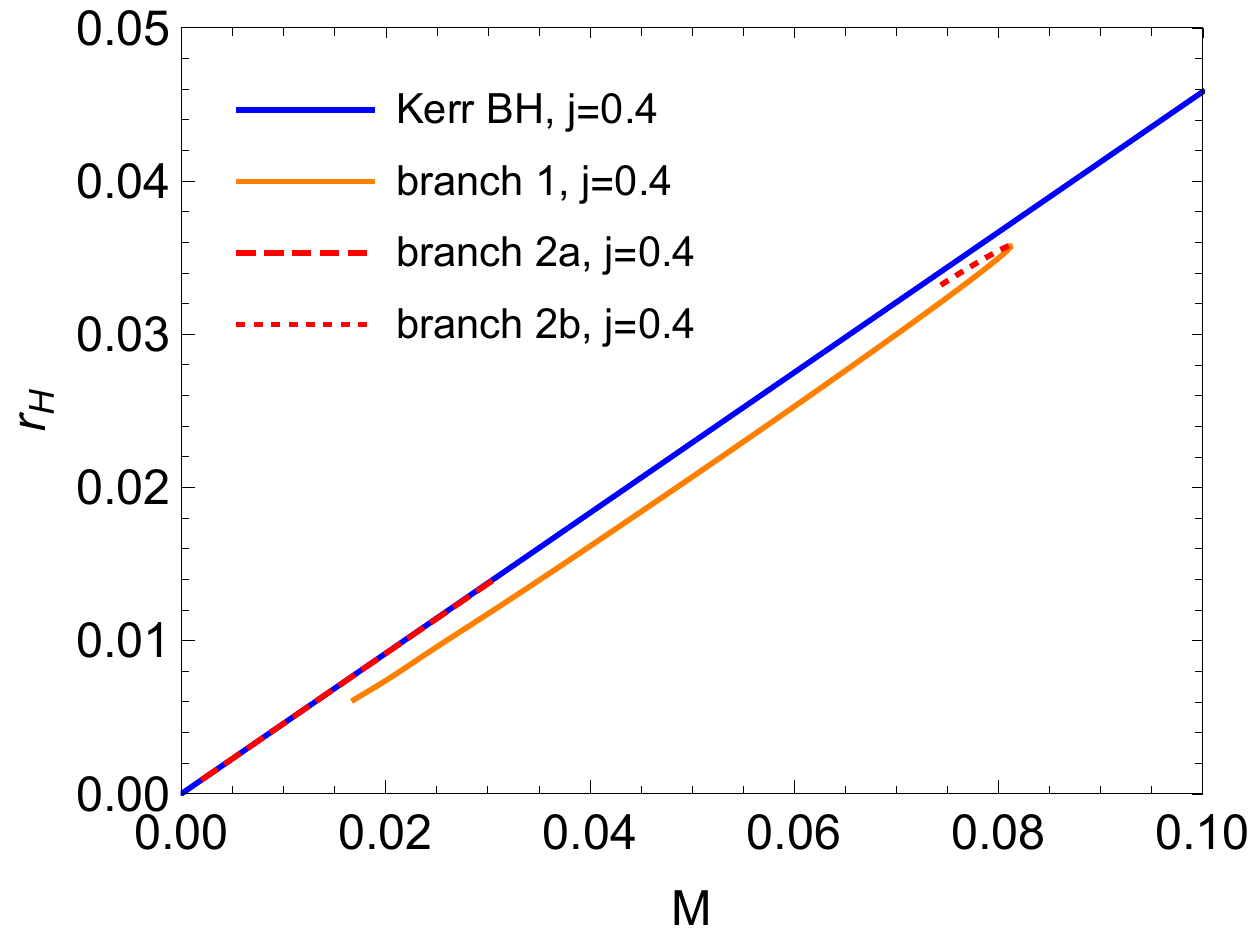}}
\caption{(Left) Scalar charge $Q_s$ as functions of the black hole mass $M$ for several different $j$. Here the black star denotes the scalarized black hole solution shown in Figs.~\ref{fig33} and \ref{fig44}.(Right) Comparison of the horizon radius $r_H$ of the scalarized rotating black holes and the Kerr Black holes with same $j=0.4$.} \label{fig22}
\end{figure*}

Considering $\kappa=400$, we plot the scalar charge $Q_s$ as a function of mass $M$ for different values of spin parameter $j$ in Figure \ref{fig22}(a). We observe that the higher the rotation parameter $j$ is, the smaller the range in $M$ is.  This contrasts to the result obtained in the decoupling limit~\cite{Doneva:2022yqu}, which states that  the larger the rotation parameter $j$ is, the larger the range in $M$ is for the same coupling function. Moreover, we discuss the $j$-dependence of $Q_s$ and $r_H$. The Figure \ref{fig22}(b) illustrates the comparison of the horizon radius $r_H$ of the scalarized rotating black holes and the Kerr Black holes with same $j=0.4$. We find that (a) the horizon radius of the scalarized black hole is smaller than that of the Kerr black hole with same $M$ and $j$, (b) branch 1 and branch 2a are connected at finite mass $M=0.0811$ and (c) branch 2b
 and Kerr BH come closer together as mass decreases.

\begin{figure}[htbp]
\centering
\subfigure{
\begin{minipage}[b]{.4\linewidth}
\centering
\includegraphics[width=0.8\textwidth]{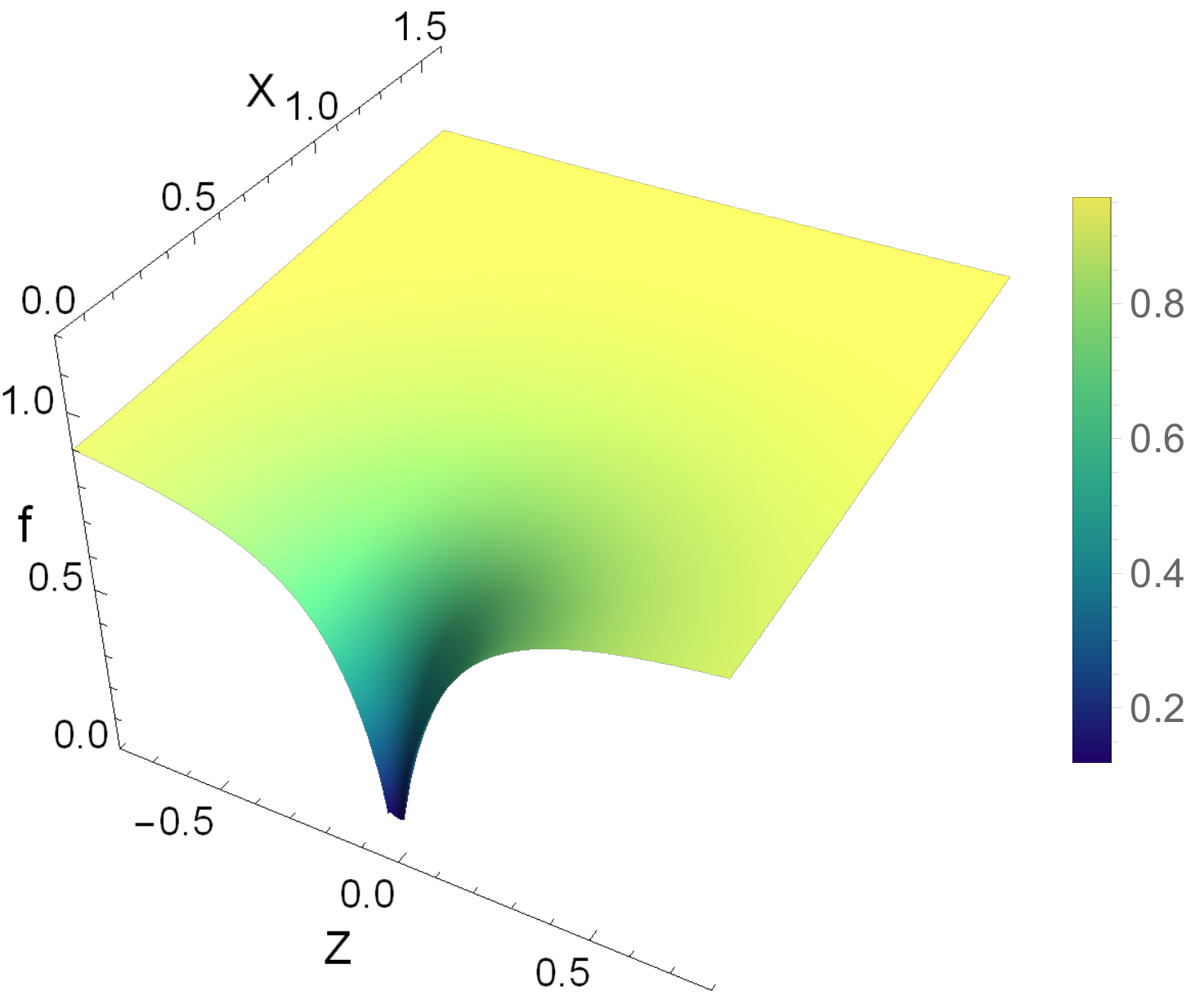}
\end{minipage}
}
\subfigure{
\begin{minipage}[b]{.45\linewidth}
\centering
\includegraphics[width=0.8\textwidth]{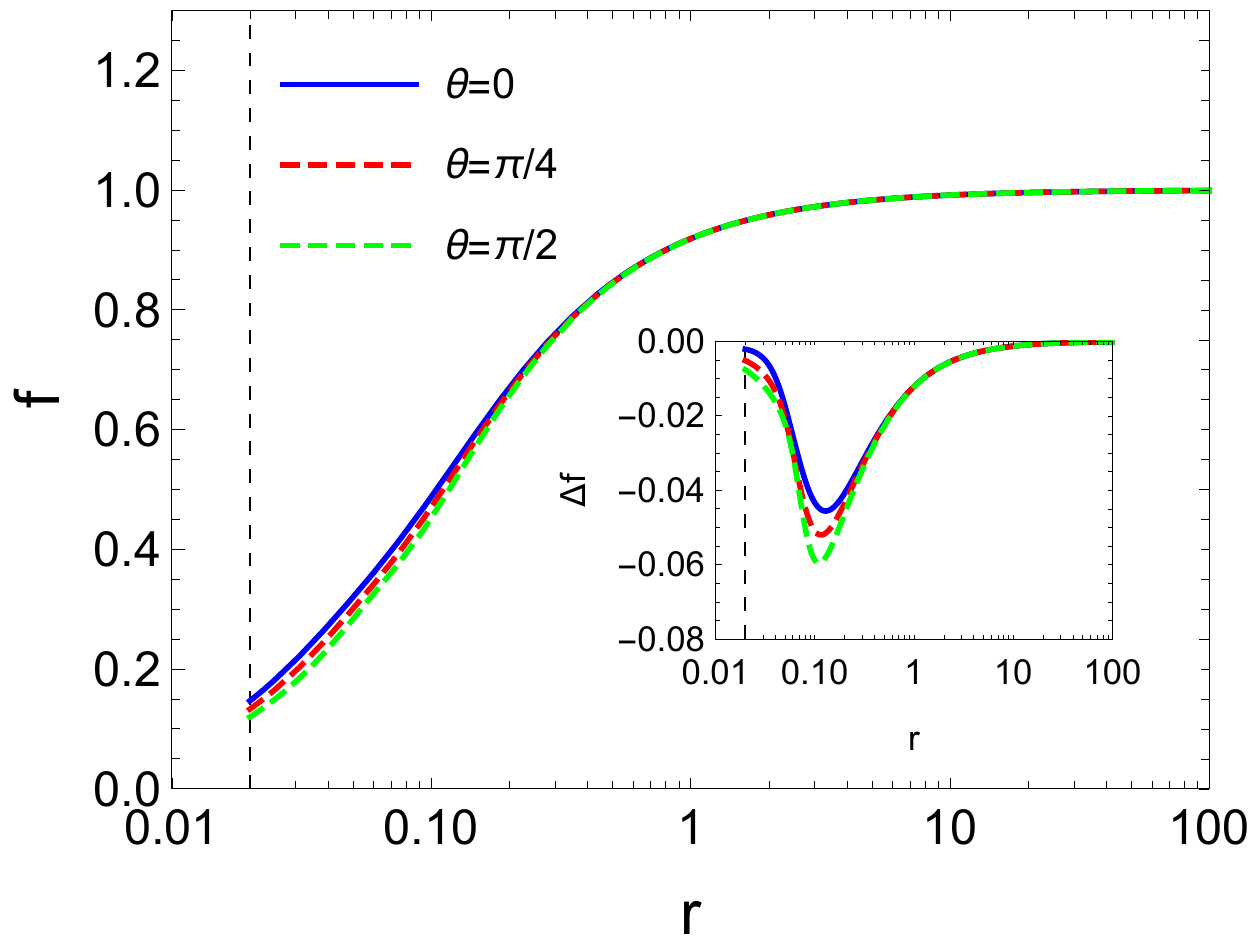}
\end{minipage}
}
\subfigure{
\begin{minipage}[b]{.4\linewidth}
\centering
\includegraphics[width=0.8\textwidth]{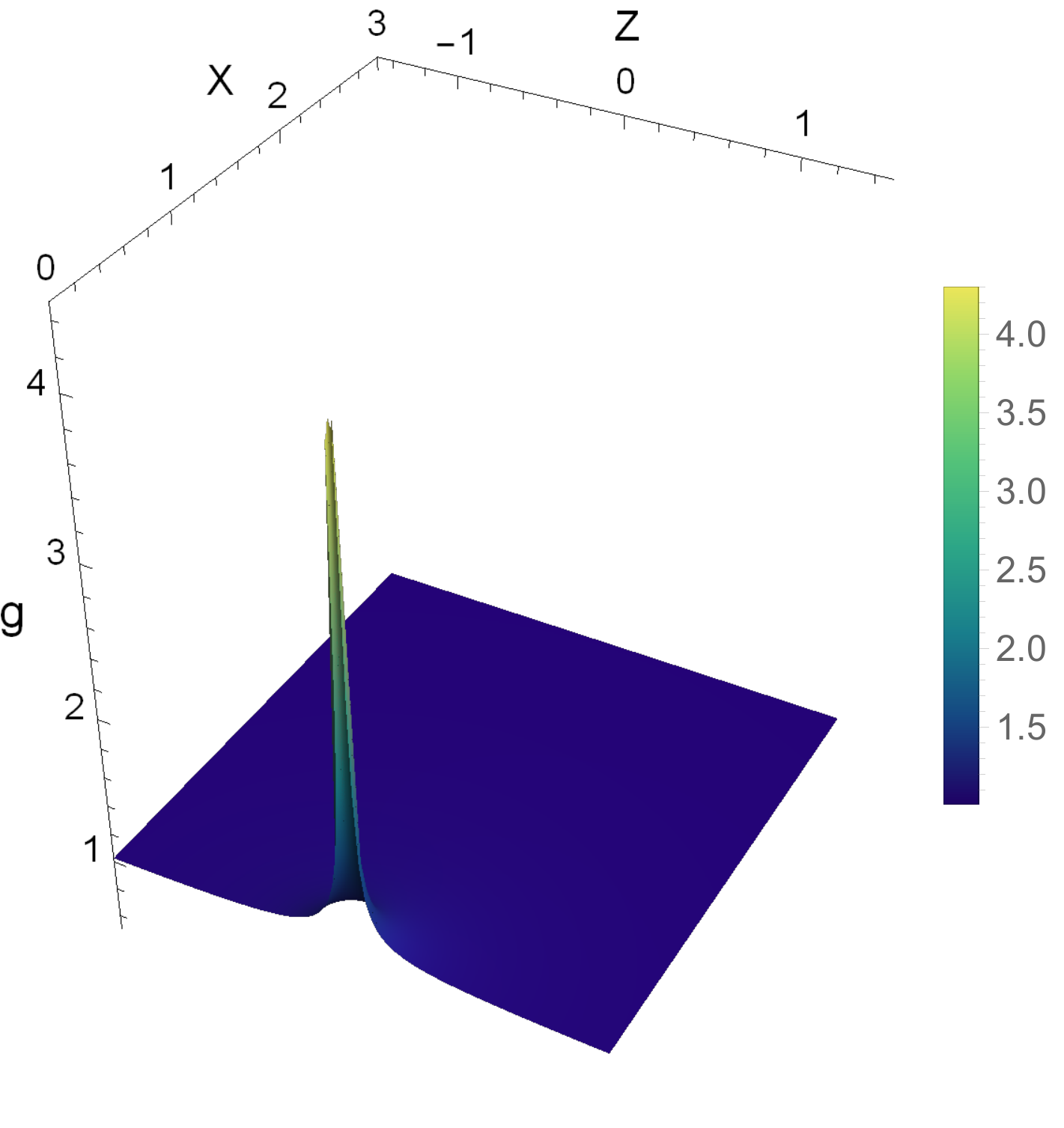}
\end{minipage}
}
\subfigure{
\begin{minipage}[b]{.45\linewidth}
\centering
\includegraphics[width=0.8\textwidth]{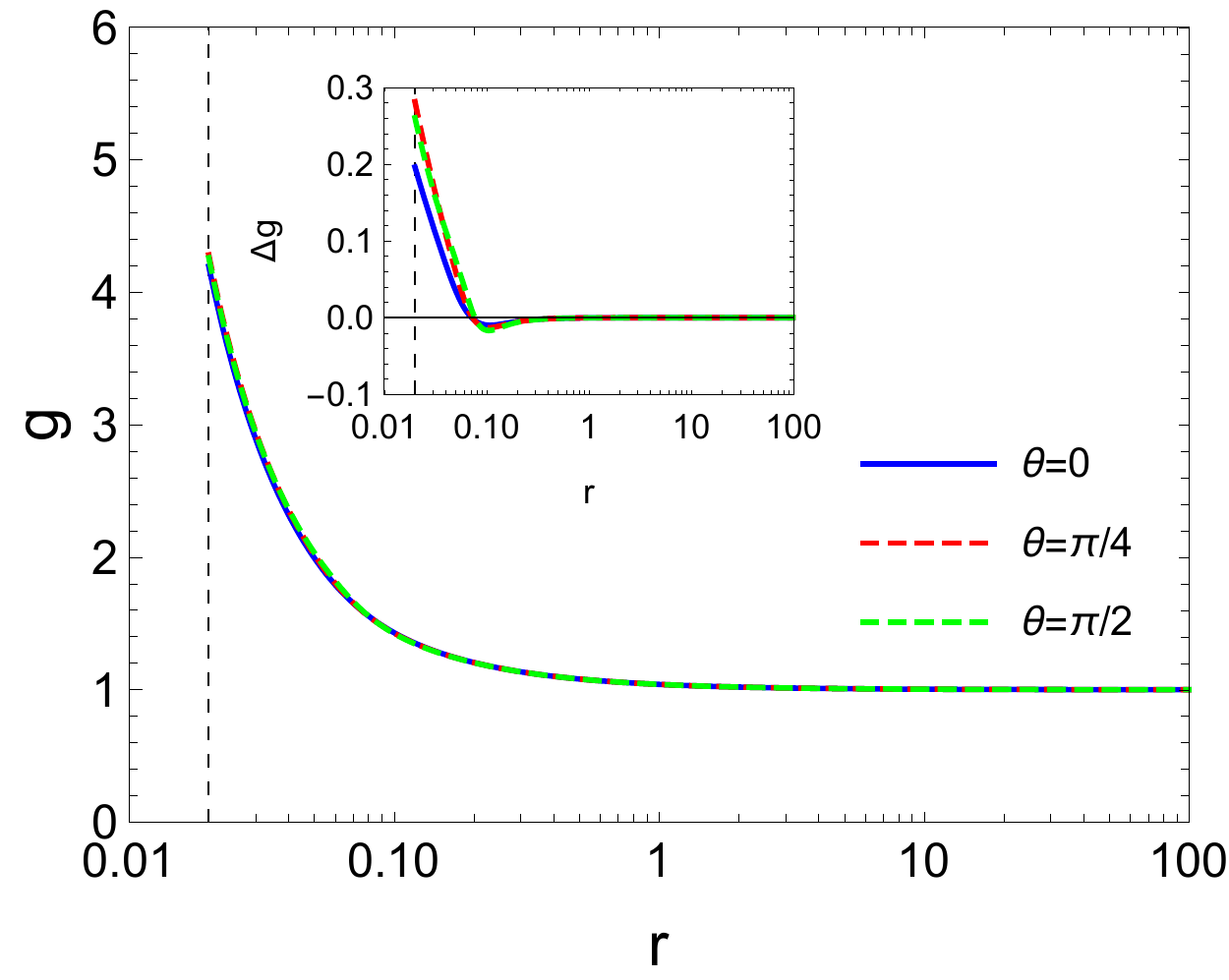}
\end{minipage}
}
\subfigure{
\begin{minipage}[b]{.4\linewidth}
\centering
\includegraphics[width=0.8\textwidth]{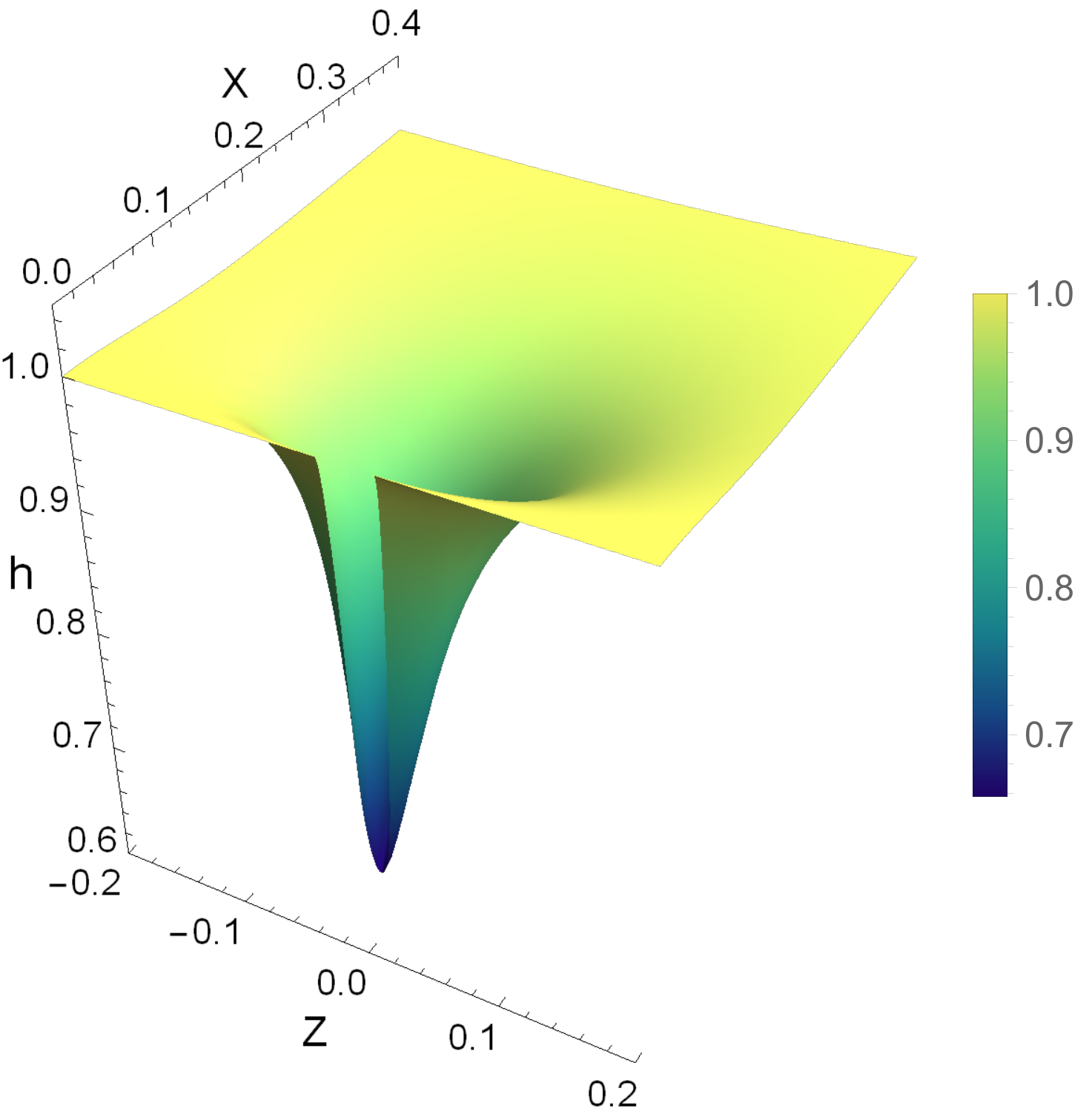}
\end{minipage}
}
\subfigure{
\begin{minipage}[b]{.45\linewidth}
\centering
\includegraphics[width=0.8\textwidth]{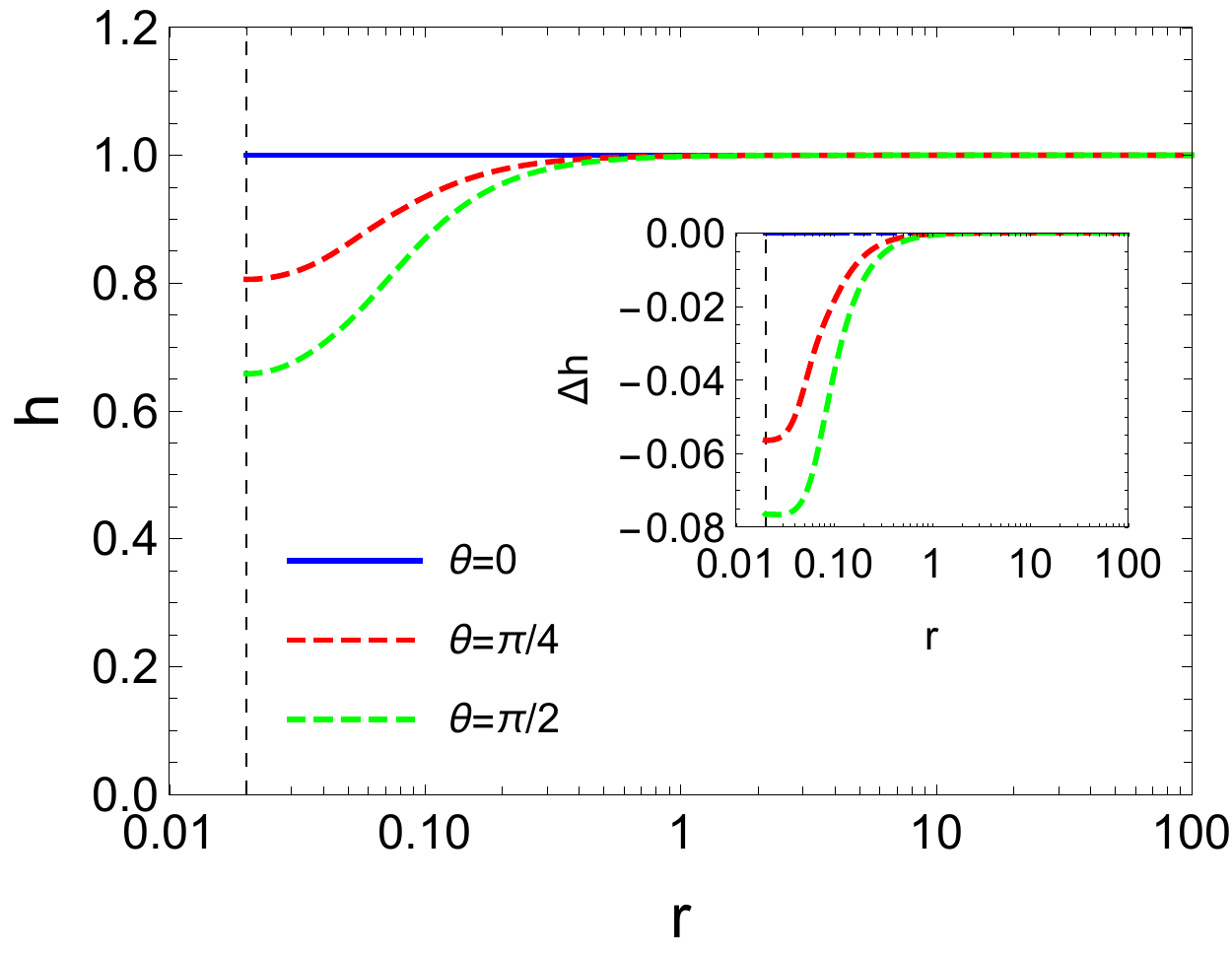}
\end{minipage}
}
\caption{Metric functions $f$, $g$ and $h$ for scalarized rotating black hole solutions with the parameters $j=a/M=0.7$, $r_H=0.02$ (dotted black line) and $\kappa=400$. The deviations between the scalarized rotating black hole and the Kerr black hole are described by $\Delta f=f-f_{kerr}$, $\Delta g=g-g_{kerr}$ and $\Delta h=h-h_{kerr}$. (Left) 3D graphs and (Right) 2D graphs.}\label{fig33}
\end{figure}

\begin{figure}[htbp]
\centering
\subfigure{
\begin{minipage}[b]{.4\linewidth}
\centering
\includegraphics[width=0.8\textwidth]{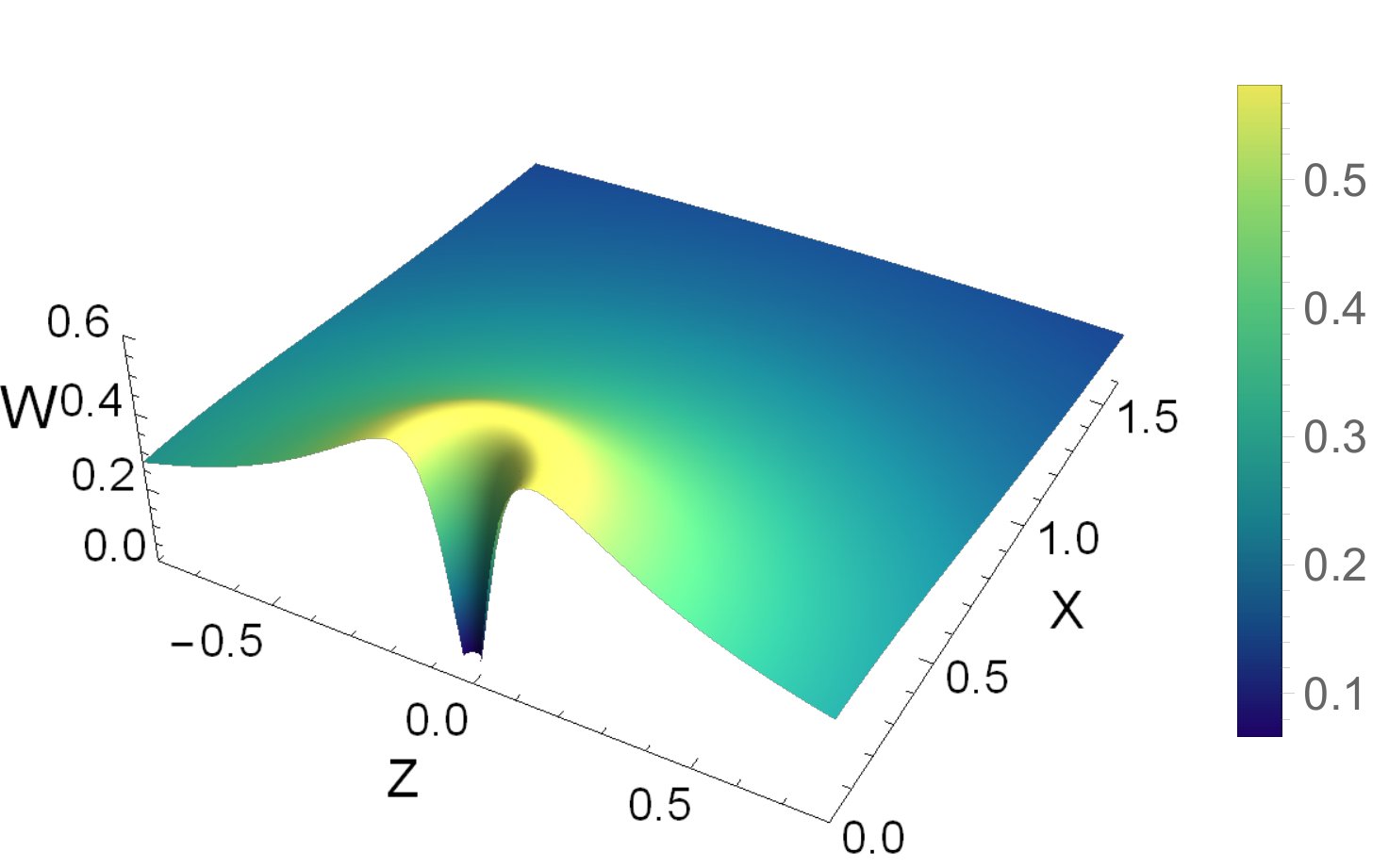}
\end{minipage}
}
\subfigure{
\begin{minipage}[b]{.45\linewidth}
\centering
\includegraphics[width=0.8\textwidth]{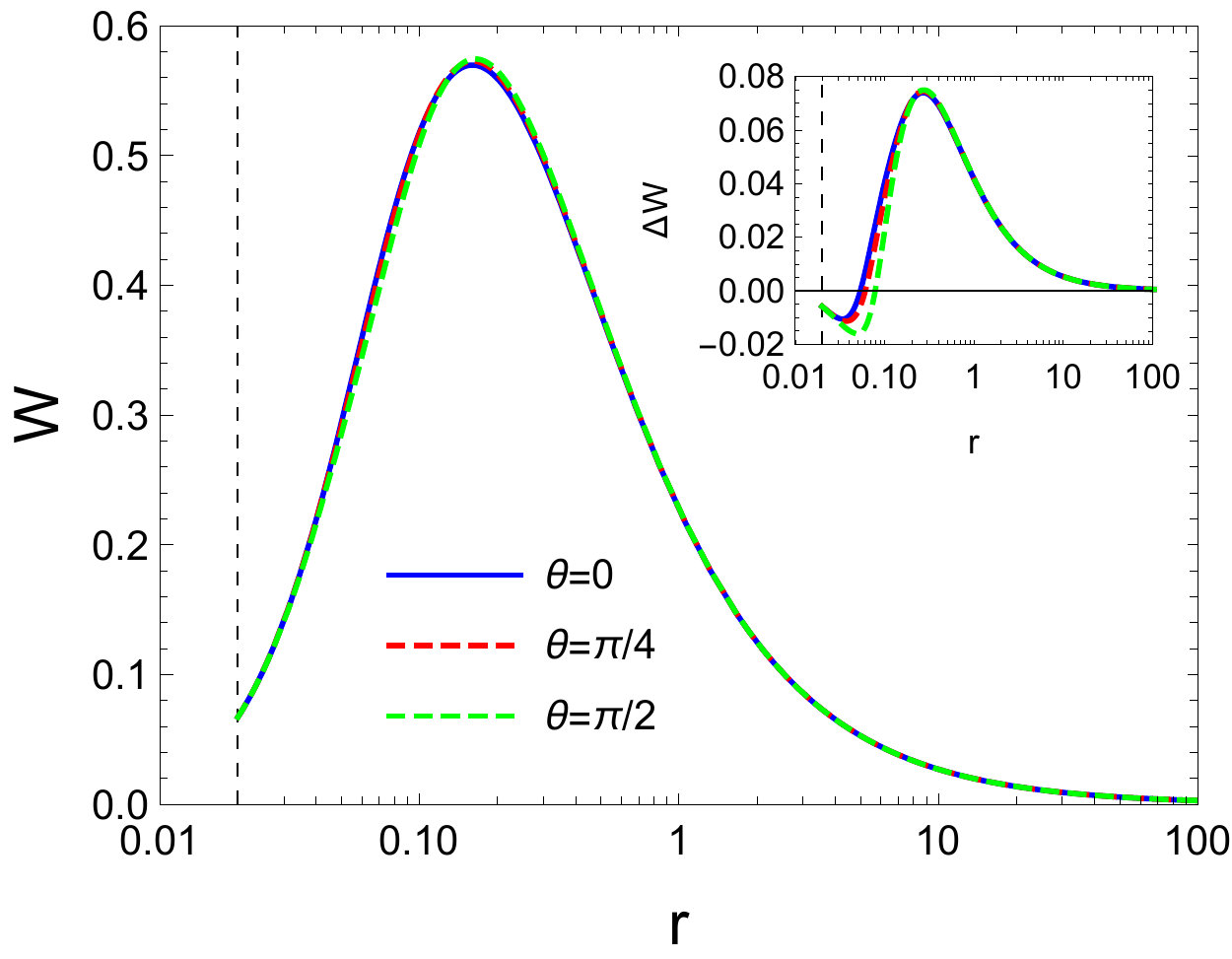}
\end{minipage}
}
\subfigure{
\begin{minipage}[b]{.4\linewidth}
\centering
\includegraphics[width=0.8\textwidth]{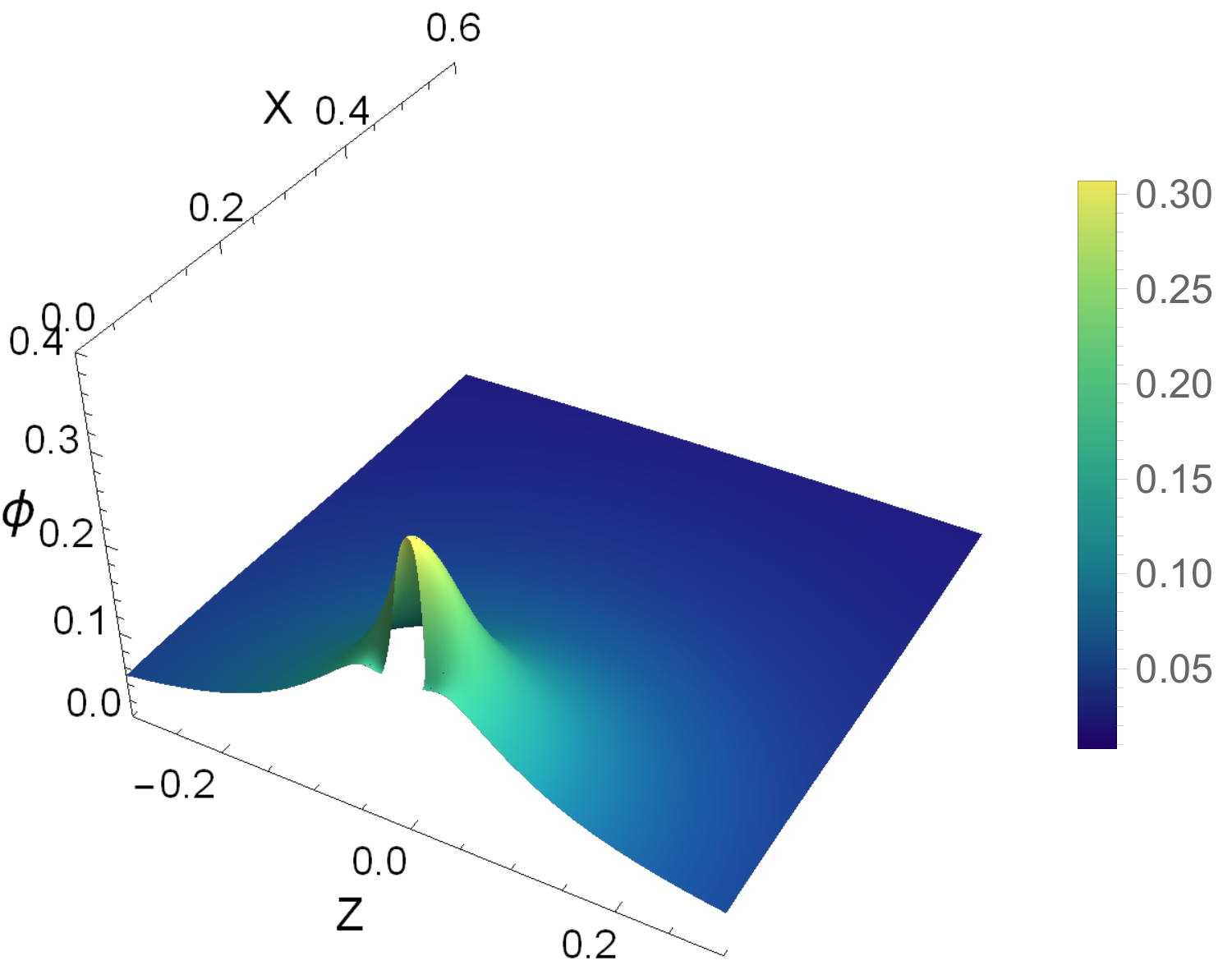}
\end{minipage}
}
\subfigure{
\begin{minipage}[b]{.45\linewidth}
\centering
\includegraphics[width=0.8\textwidth]{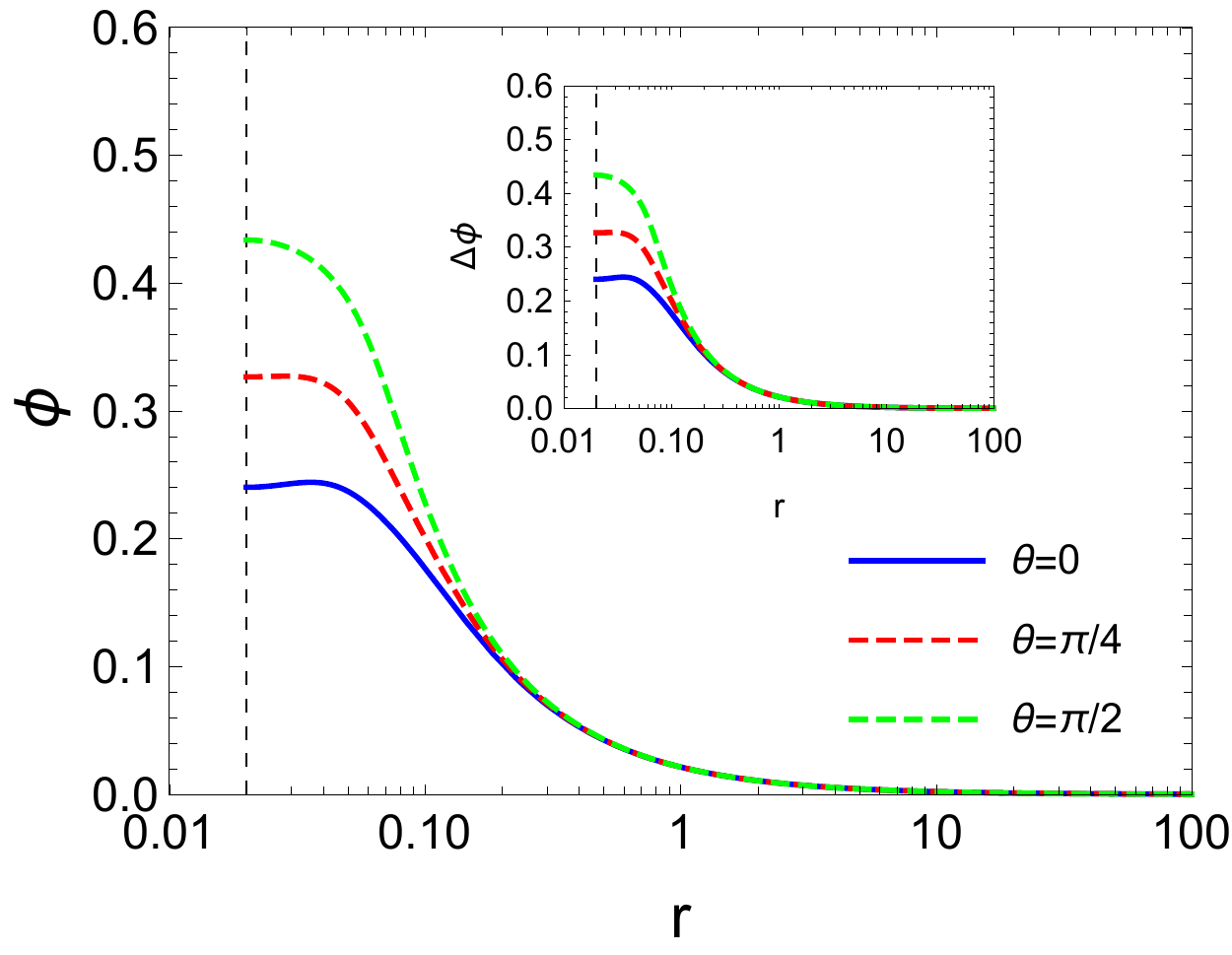}
\end{minipage}
}
\caption{Metric function $W$ and scalar field $\phi$  represent  the nonlinearly scalarized rotating black hole solution
with the same parameters as in Fig.~\ref{fig33}. The solution deviations between the scalarized rotating black
hole and the Kerr black hole are described by $\Delta W=W-W_{kerr}$ and $\Delta \phi=\phi-\phi_{kerr}$. (Left) 3D graphs and (Right) 2D graphs. }\label{fig44}
\end{figure}

Choosing the parameters $j=a/M=0.7$, $r_H=0.02$ of black hole, we present the newly scalarized rotating black hole solutions with $\kappa=400$ in Figures \ref{fig33} and \ref{fig44}, where the left column shows 3D plots and right column displays 2D plots of the corresponding functions in terms of the radial variable for three different values of the angular coordinate. Here,  the axes for the 3D plots are $X=r \sin\theta$ and $Z=r\cos\theta$ (with $r\geq r_H$). With the horizon radius $r_H=0.02$, a scalarized rotating black hole with $M=0.0626$, $j\equiv J/M^2=0.7$ and $Q_s=0.0152$ is obtained numerically. Holding the same $r_H$ and $j$, the solution deviations between the scalarized rotating black hole and Kerr black hole are also presented in these figures. We observe from  Figures \ref{fig33} and \ref{fig44} that our nonlinearly scalarized rotating black hole solution represents  an asymptotically flat rotating black hole with scalar hair and clear $\theta$-dependence for $h$ and $\phi$.

\section{THERMODYNAMIC PROPERTIES} \label{sec4}

\begin{figure}[htbp]
\subfigure{
\begin{minipage}[b]{.45\linewidth}
\centering
\includegraphics[width=0.85\textwidth]{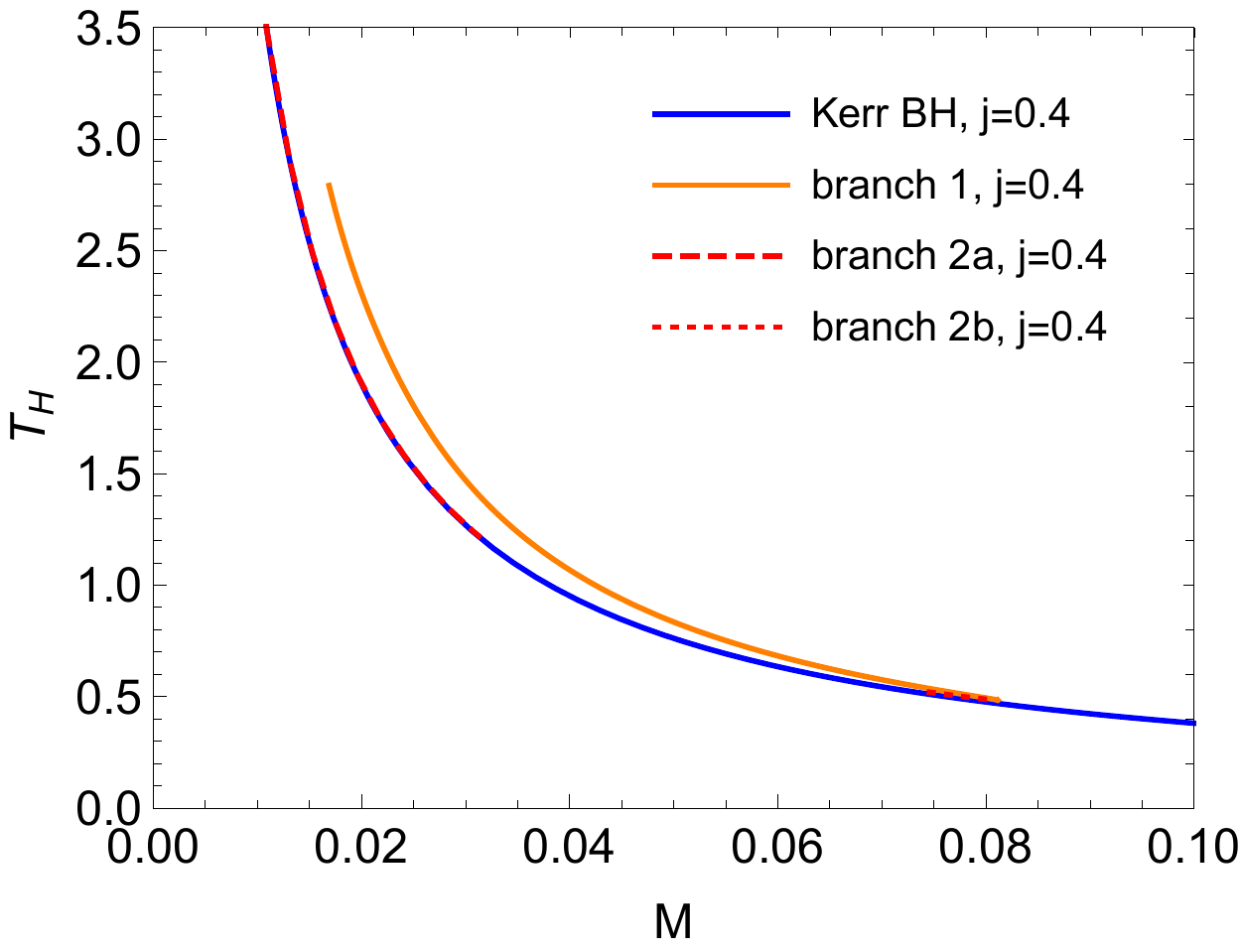}
\end{minipage}
}
\subfigure{
\begin{minipage}[b]{.45\linewidth}
\centering
\includegraphics[width=0.85\textwidth]{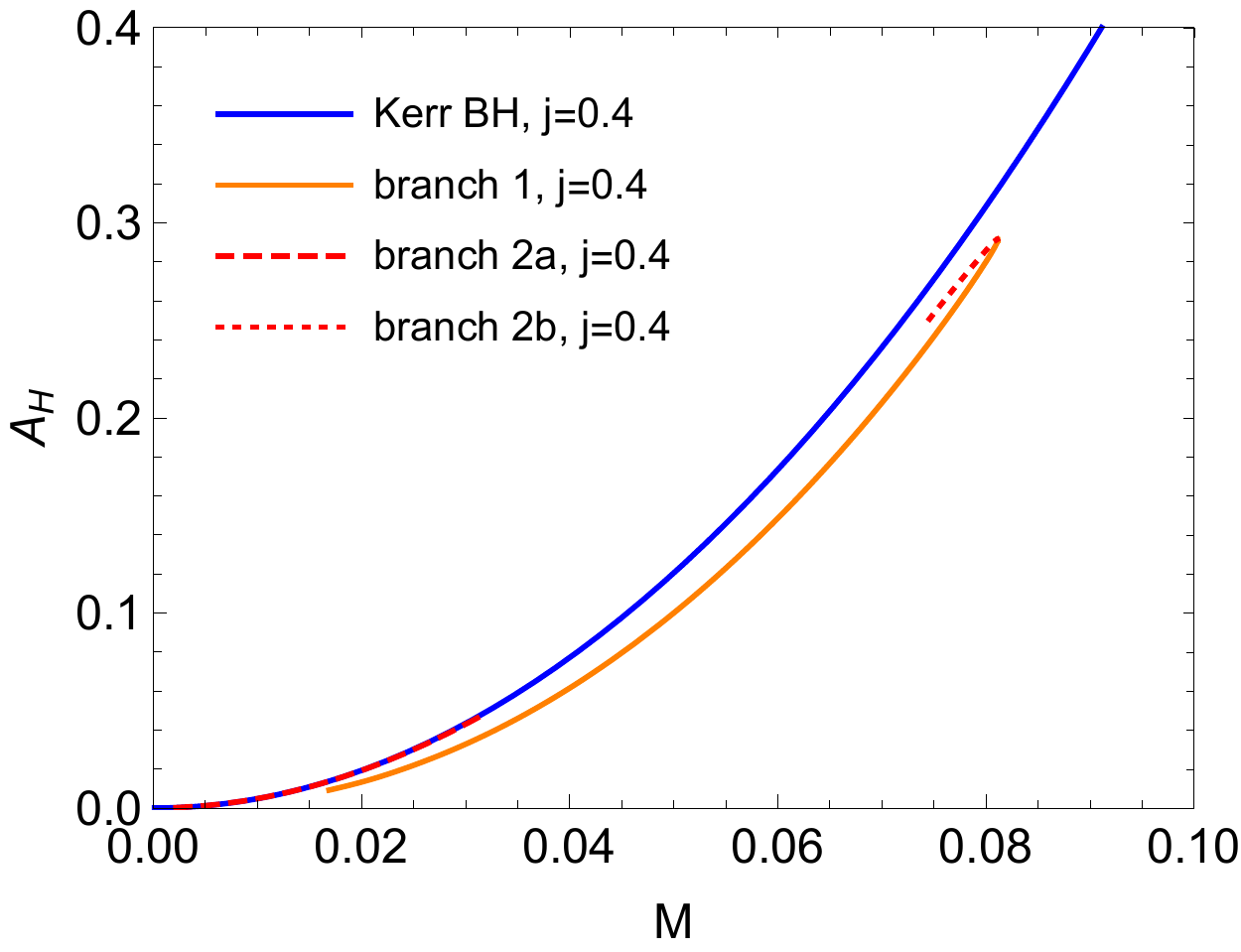}
\end{minipage}
}
\caption{(Left) Comparison of the Hawking temperature $T_H$ of the scalarized rotating black holes and Kerr Black holes with same $j=0.4$. (Right) Comparison of the horizon area $A_H$ of the scalarized rotating black holes and Kerr Black holes with same $j=0.4$.}\label{fig5}
\end{figure}

Now, we further discuss the thermodynamic properties of scalarized rotating black holes.
The surface gravity is defined as $\zeta^2=-\frac{1}{2}(\nabla_\mu\chi_\nu)(\nabla^\mu\chi^\nu)$.
Then, the Hawking temperature of black holes takes the form
\begin{eqnarray}
T_H=\frac{\zeta}{2\pi}=\frac{1}{2\pi r_H}\frac{f}{\sqrt{gh}}|_{x=-1}.
\end{eqnarray}
The comparison of the Hawking temperature $T_H$ of the scalarized rotating black holes and the Kerr Black holes with same $j=0.4$ is shown in the left panel of Fig. \ref{fig5}. The Hawking temperature of the scalarized black hole is higher than that of the Kerr black hole with same $M$. Moreover, the stationary and rotational symmetry metric \eqref{metric} possesses two Killing vector fields
\begin{eqnarray}
\xi=\partial_t, \quad \eta=\partial_\varphi
\end{eqnarray}
and its  linear combination
\begin{eqnarray}
\chi=\xi+\Omega_{H}\eta,
\end{eqnarray}
where the angular velocity $\Omega_{H}$ is determined by the horizon value of the metric function
\begin{eqnarray}
\Omega_{H}=-\frac{\xi\cdot\eta}{\eta\cdot\eta}=-\frac{g_{\varphi t}}{g_{\varphi \varphi}}|_{x=-1}=\frac1{r_H}W|_{x=-1}.
\end{eqnarray}

Let us compute the entropy of scalarized rotating black holes. In the
EsGB gravity, the black hole entropy is not given by the Bekenstein-Hawking formula. Concerning the horizon properties, we note that the induced metric on the horizon is given by
\begin{eqnarray}
d\Sigma^2=\gamma_{ij}dx^i dx^j=r_H^2\frac{g}{f}\left(hd\theta^2+\sin^2\theta d\varphi^2\right)|_{x=-1}.
\end{eqnarray}
The horizon area is obtained as
\begin{eqnarray}
A_H=\int_H \sqrt{\gamma}d\theta d\varphi=2\pi r_H^2\int^{\pi}_{0}d\theta\sin\theta\frac{g\sqrt{h}}{f}|_{x=-1}.
\end{eqnarray}
We plot the the comparison of the horizon area $A_H$ of the scalarized rotating black holes and Kerr Black holes with same $j=0.4$ in the right panel of Fig. \ref{fig5}. The horizon area of the scalarized black hole is smaller than that of Kerr black hole with same $M$ and the branches of the scalarized black hole have similar behavior as in the left panel of Fig. \ref{fig44} and Fig. \ref{fig5}.

\begin{figure}[htbp]
\centering
\includegraphics[width=0.4\textwidth]{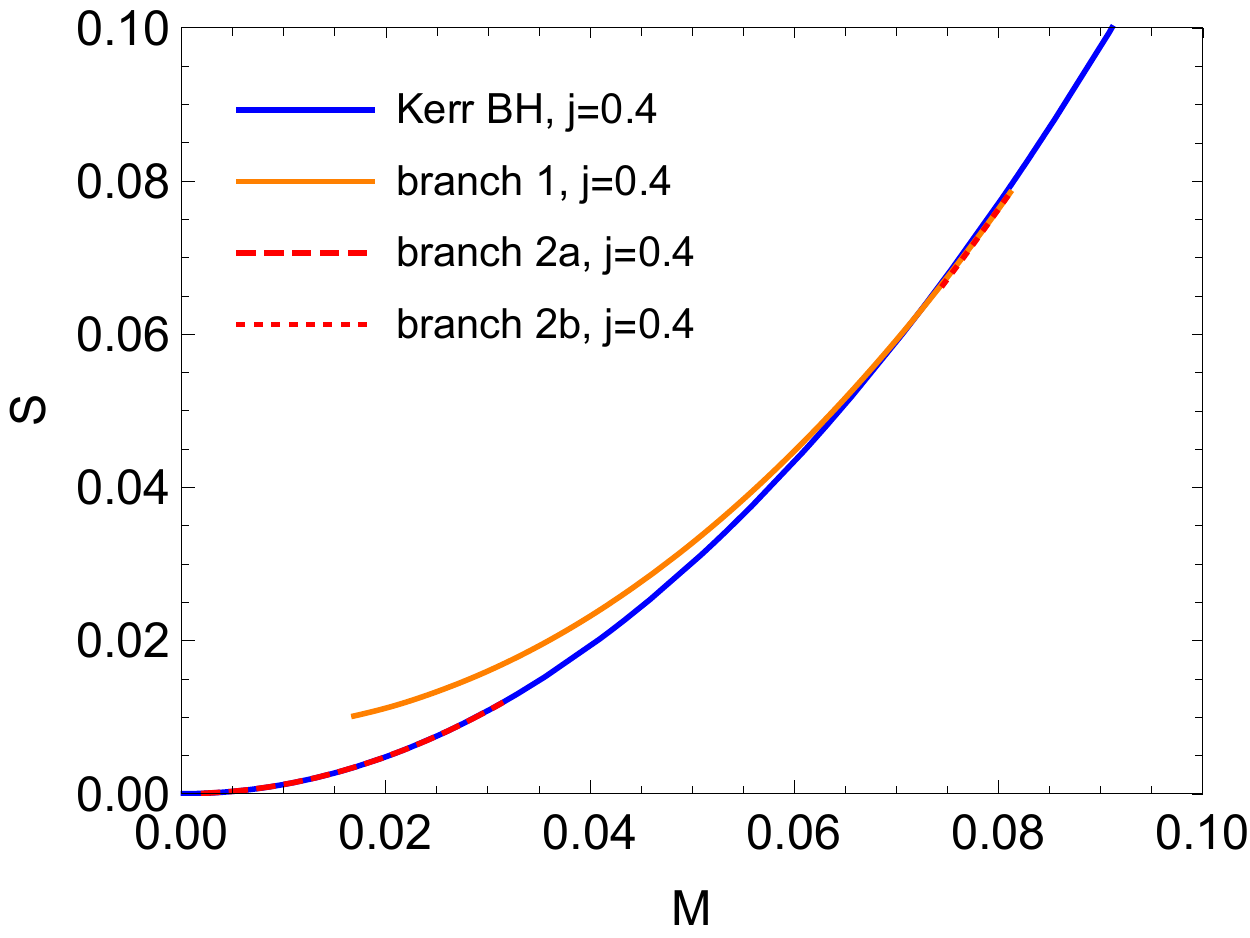}
\caption{Comparison of the entropy $S$ of the scalarized rotating black holes and Kerr Black holes with same $j=0.4$.}\label{fig6}
\end{figure}

Then, the entropy defined by the Iyer-Wald formalism is
\begin{eqnarray}
S=-2\pi\int_H \frac{\delta {\cal L}}{\delta R_{\mu\nu\alpha\beta}}\epsilon_{\mu\nu}\epsilon_{\alpha\beta}dA|_{\rm on-shell},
\end{eqnarray}
where $\epsilon_{\mu\nu}$ is the binormal vector to the horizon surface.
In Fig.~\ref{fig6}, we plot the the comparison of the entropy $S$ of the scalarized rotating black holes and Kerr Black holes with same $j=0.4$. The curves branch 1 and Kerr black hole have an intersection at finite mass and the entropies of branch 2a and 2b are always smaller than that of branch 1 with same $M$. This suggest that the entropic preference of branch 1 and Kerr black hole undergo a shift, while branch 1 is always entropically favored over branch 2a and 2b. Moreover, the branches of the scalarized black hole have similar behavior as in the previous figures. 

Based on these physical quantities of the scalarized rotating black holes, we can  further check the Smarr relation. Actually, the Smarr relation plays an important role  when studying numerical solutions, since it provides a testbed to the code that relates physical quantities obtained on the horizon to those obtained  asymptotic regions, and also allows us to estimate the accuracy of our numerical method. This relation is given by
\begin{eqnarray}
M+M_s=2T_H S+2\Omega_H J,
\end{eqnarray}
where $M_s$ is a bulk (outside the horizon) integral along a spacelike hypersurface $\Sigma$ \cite{Fernandes:2022gde,Smarr:1972kt,Liberati:2015xcp}
\begin{eqnarray}
M_s=-\frac{1}{2\pi}\int d^3 x\sqrt{-g}\frac{F(\phi)}{F'(\phi)}\square{\phi}.
\end{eqnarray}
and can be related to the scalar charge $Q_s$ of scalarized rotating  black holes.
 We present several discrete values of these thermodynamic quantities $M_s$, $J$, $T_H$, $S$ and $\Omega_H$ for scalarized rotating black holes to test Smarr relation, as shown in Table 1.
 We check that  these thermodynamic quantities obey the Smarr formula with high precision.

\begin{table}[h]\label{table1}
\centering
{\begin{tabular}{|c|c|c|c|c|c|c|c|}
\hline
No. & $j$  & $M_s$    & $T_H$  & $S$      & $\Omega_H$ & $J$ & Smarr \\ \hline
1   &0 & 0.00239   & 0.506        & 0.0833  & 0          &0 & $5.23\times10^{-6}$ \\ \hline
2   &0.0485 & 0.00238   & 0.505        & 0.0832  & 0.157          &$3.25\times10^{-4}$ & $5.26\times10^{-6}$\\ \hline
3   &0.118 & 0.00233   & 0.503        & 0.0830  & 0.380         &$7.87\times10^{-4}$ & $5.07\times10^{-6}$\\ \hline
4   &0.179 & 0.00226   & 0.501        & 0.0826  & 0.582          & 0.00120 & $5.08\times10^{-6}$\\ \hline
5   & 0.235 & 0.00215   & 0.497        & 0.0821  & 0.765          & 0.00157 & $4.83\times10^{-6}$\\ \hline
6   & 0.289 & 0.00197   & 0.492        & 0.0814  & 0.943         & 0.00193 & $4.53\times10^{-6}$\\ \hline
\end{tabular}}
\caption{ Six discrete values of thermodynamic quantities $M_s$, $J$, $T_H$, $S$ and $\Omega_H$ for scalarized rotating black holes are displayed  for testing the Smarr relation with fixed mass $M=0.0818$.}
\end{table}

\begin{figure*}[t!]
\centering
\includegraphics[width=0.6\textwidth]{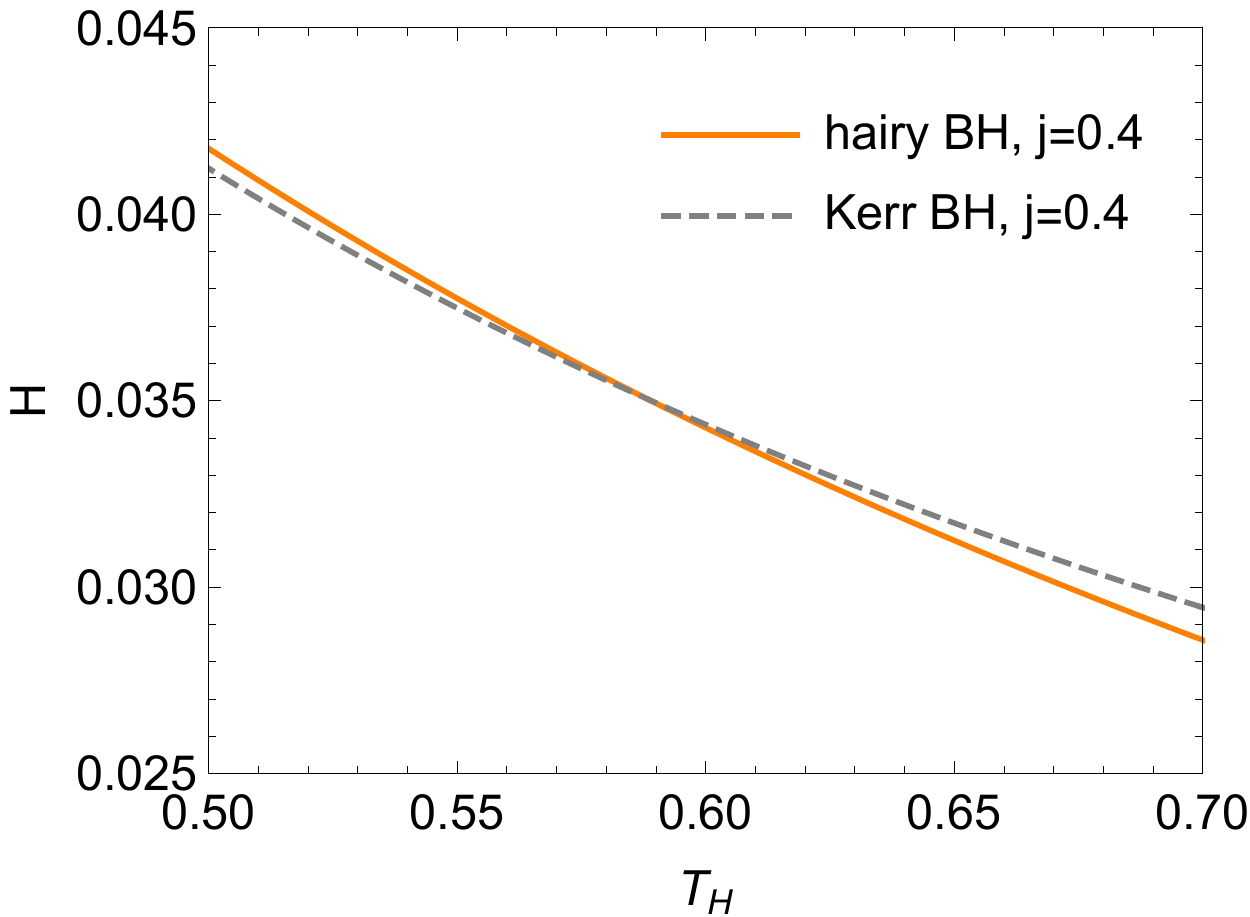}
\caption{Free energy $H$ versus Hawking temperature $T_H$ curves for scalarized rotating
and Kerr black holes with $j=0.4$ and $\kappa=400$.}\label{fig7}
\end{figure*}

 Finally, we note that the Helmholtz (on-shell) free energy $H=M-T_H S$ as a function of temperature is important to check a phase transition between scalarized rotating and Kerr black holes in canonical ensemble~\cite{Myung:2008ze}. We shown in Fig. \ref{fig7} with fixed $j=0.4$ when choosing the temperature matching. We see that the free energy of Kerr black hole without scalar hair is always lower than that of scalarized rotating black hole for  $T_{H}<T_c$, it crosses  the critical point at $T_H=T_c\approx 0.589$, and then, it becomes higher than that for scalarized rotating black hole for $T_{H}>T_c$. In other words,  for $T_{H}<T_c$, the Kerr black hole is more favorable than the scalarized rotating black hole, while for $T_{H}>T_c$, the scalarized rotating black hole is thermodynamically favorable than the Kerr black hole. This means that for $T_{H}<T_c$, the ground state is Kerr black hole, whereas for $T_{H}>T_c$, the ground state is given by the scalarized rotating black hole. It implies that a first order phase transition  may occur between Kerr and scalarized rotating black holes in the EsGB gravity.  Moreover, we note that the critical temperature $T_c$ increases with increasing spin $j$ (see Table 2). In particular, when $j\gtrsim 0.69$, the temperature of the black hole is always less than the critical temperature.

\begin{table}[h]\label{table2}
\centering
{\begin{tabular}{|c|c|c|c|c|c|c|c|}
\hline
$j$ &     0      & 0.1     & 0.2    & 0.3     &0.4  & 0.5 & 0.6 \\ \hline
$T_c$&  0.560  & 0.562   & 0.568  & 0.577  & 0.589   &0.600 &  0.614 \\ \hline
\end{tabular}}
\caption{ The discrete values of the critical temperature $T_c$ for different spin $j$.}
\end{table}
\section{CONCLUSIONS AND DISCUSSION} \label{sec5}

In the paper, we have discussed the nonlinear scalarization of rotating black holes in EsGB gravity, markedly different from the conventional
spontaneous scalarization for black holes.  We have considered the quartic coupling functions $F(\phi)$ in (\ref{funterm})
for which the Kerr black hole is still a linearly stable solution of the field equations but for certain ranges of the parameters,
nonlinearly scalarized phases of the Kerr black hole appear.
This is because  even though the Kerr black hole is stable against small (linear) perturbations, this linear stability can be lost
for larger amplitudes of the scalar perturbations that will bring us to the nonlinear regime.

 In order to obtain a full spectrum of scalarized rotating black holes  including the unstable ones, we have solved the fully nonlinear
coupled system of reduced field equations by using the pseudospectral method. We have obtained multibranches of scalarized
rotating black hole solutions for different values of parameter $\kappa$ in the coupling functions [Eq.(\ref{funterm})].
Moreover, we focused on the stable branch of scalarized rotating black holes with the specific value of $\kappa=400$, and further
studied the thermodynamic properties for these scalarized rotating black holes in detail. We found these
thermodynamic properties of these black holes obey the Smarr relation, which allows us to check the accuracy of our numerical method.
Finally,  we have investigated the phase transition of two black holes by evaluating the free energy for Kerr and scalarized rotating black holes.
It is clear that the free energy of scalarized rotating black holes is always higher than that of Kerr black holes without
scalar hair for  $T_{H}<T_c$, it crosses  the critical point at $T_H=T_c$, and then it becomes lower than that for Kerr black holes for $T_{H}>T_c$.
This implies that a first order phase transition  may occur between Kerr and scalarized rotating black holes in EsGB gravity.

As a general extension of this work, it is interesting to explore the dependence of these nonlinearly scalarized black holes
on various forms of coupling functions. In addition, it would be  better to study the effect of spin on the dynamical stability
of nonlinearly scalarized black holes.  How to relate its properties to astronomical observations is also one of important issues.
Moreover, we found that the stable scalarized phase has the largest entropy among all the branches of
hairy black holes. It also has larger entropy than the Kerr phase for most of the parameter range,
making it thermodynamically preferred. We have reasons to believe that their solutions would possibly be the endpoint of the scalarization.
The stability analysis for these scalarized rotating black holes is also a feasible way to check the conjecture in future.
These plans for the next work will contribute to a better understanding of the nonlinear scalarization mechanism.

 \vspace{1cm}

{\bf Acknowledgments}
 \vspace{1cm}

We appreciate Eugen Radu and Hyat Huang for helpful discussion. This research is supported by the National Key
Research and Development Program of China under Grant
No. 2020YFC2201400. M. Y. L is also supported by  the Jiangxi Provincial Natural Science Foundation (Grant No. 20224BAB211020) and the Science and Technology Program of Guangxi, China (Grant No. 2018AD19310).
 D. C. Z acknowledges financial support from the Initial Research Foundation of Jiangxi Normal University.
 \vspace{1cm}

\appendix
\section*{APPENDIX: NUMERICAL SCHEME}

In this appendix, we briefly present the calculation details of numerical solutions of rotating scalarized black holes by using the pseudospectral and Newton-Raphson methods.

Considering a compactified radial coordinate and symmetries of our problem,
a suitable spectral expansions for four black hole metric functions and scalar field (collectively denoted
by ${F}={f,g,h,W,\phi}$) are given by
\begin{eqnarray}
{F}^{(k)} =\sum_{i=0}^{N_x-1}\sum_{j=0}^{N_\theta-1}\alpha_{ij}^{(k)}T_i(x)\cos(2j\theta), \label{spexpansion}
\end{eqnarray}
where $N_x$ and $N_\theta$ denote the resolutions in the radial and angular coordinates. We note that the angular boundary conditions are automatically satisfied by this
expansion.

As mentioned previously, we will use the Kerr metric itself to set our
initial guess when working with EsGB gravity. Thus,  we
need  the spectral coefficients  expressed in terms of  interpolation of
a two-dimensional function $u(x,\theta)$
\begin{eqnarray} \label{alpha}
\alpha_{ij}=\frac{4}{N_x N_\theta}\sum^{N_x-1}_{k=0}\sum^{N_{\theta}-1}_{l=0}u(x_k,x_\theta)T_i(x_k)\cos(2j\theta_l),
\end{eqnarray}
where $x_k$ and $\theta_l$ are given by
\begin{eqnarray}
x_k=\cos\Big[\frac{(2k+1)\pi}{2N}\Big],\quad
\theta_l=\frac{(2l+1)\pi}{4N}, \quad k,l=0,..., N-1,
\end{eqnarray}
respectively.
\begin{figure}
\centering
\includegraphics[width=0.55\textwidth]{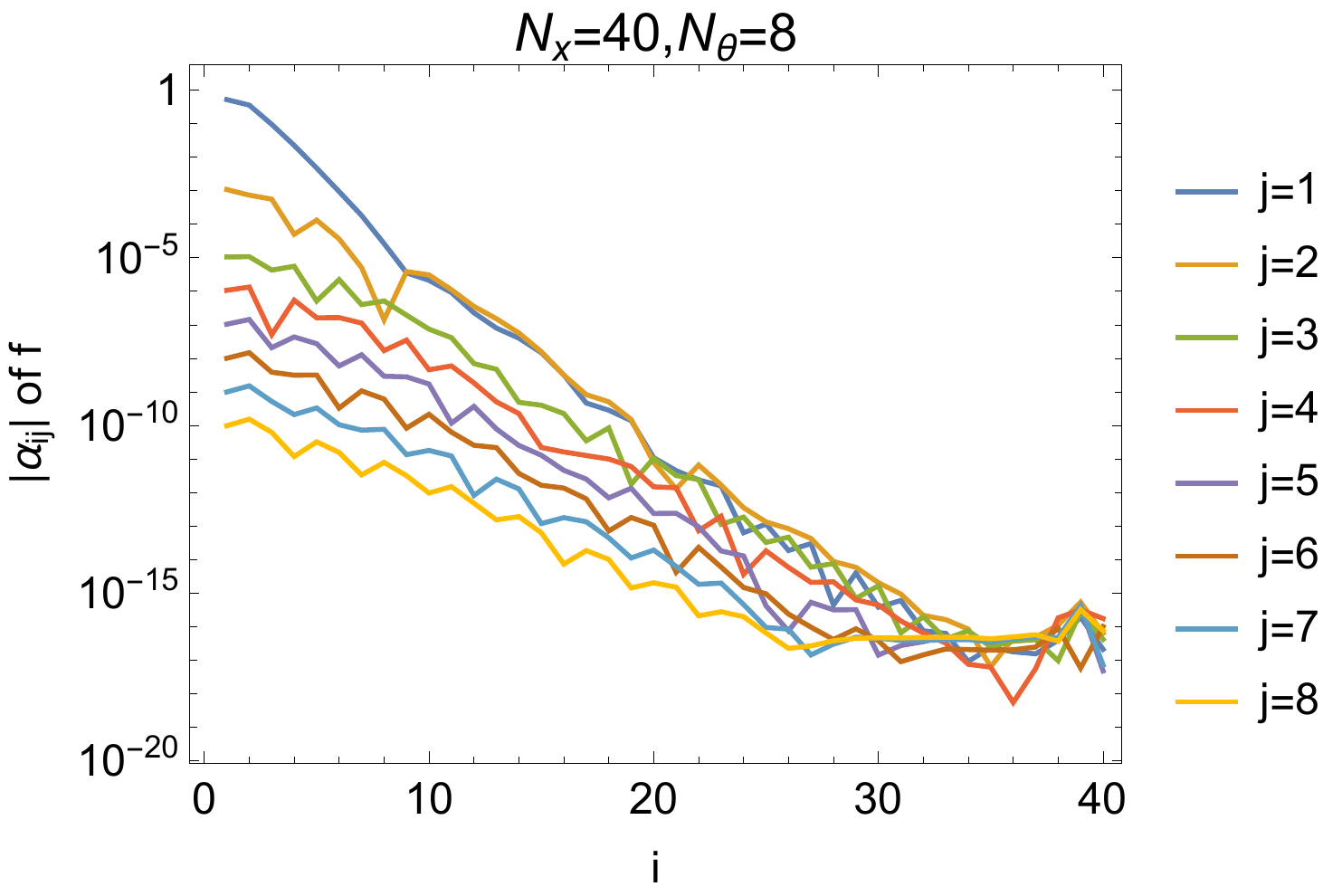}
\hfill%
\includegraphics[width=0.55\textwidth]{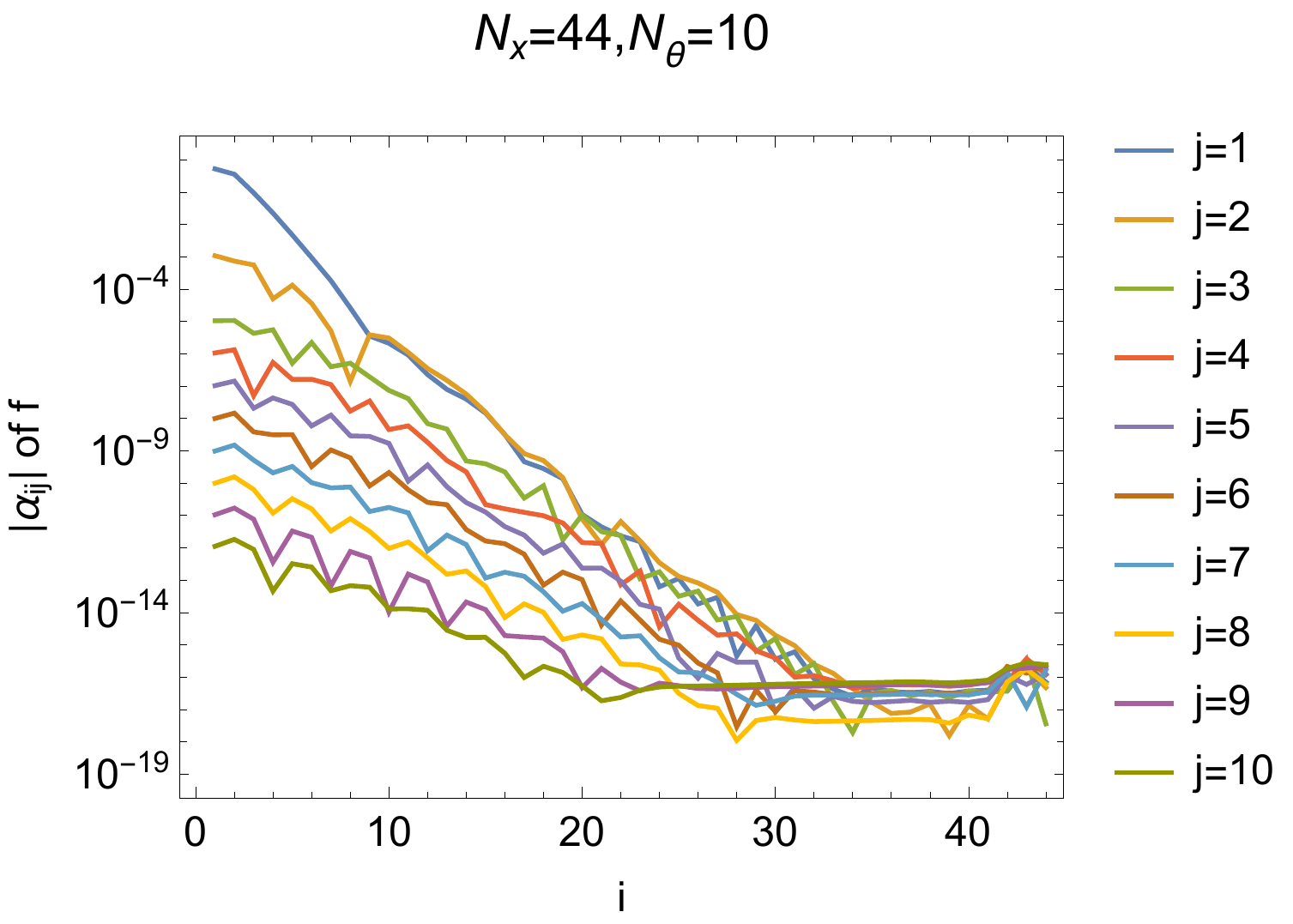}
\hfill%
\includegraphics[width=0.65\textwidth]{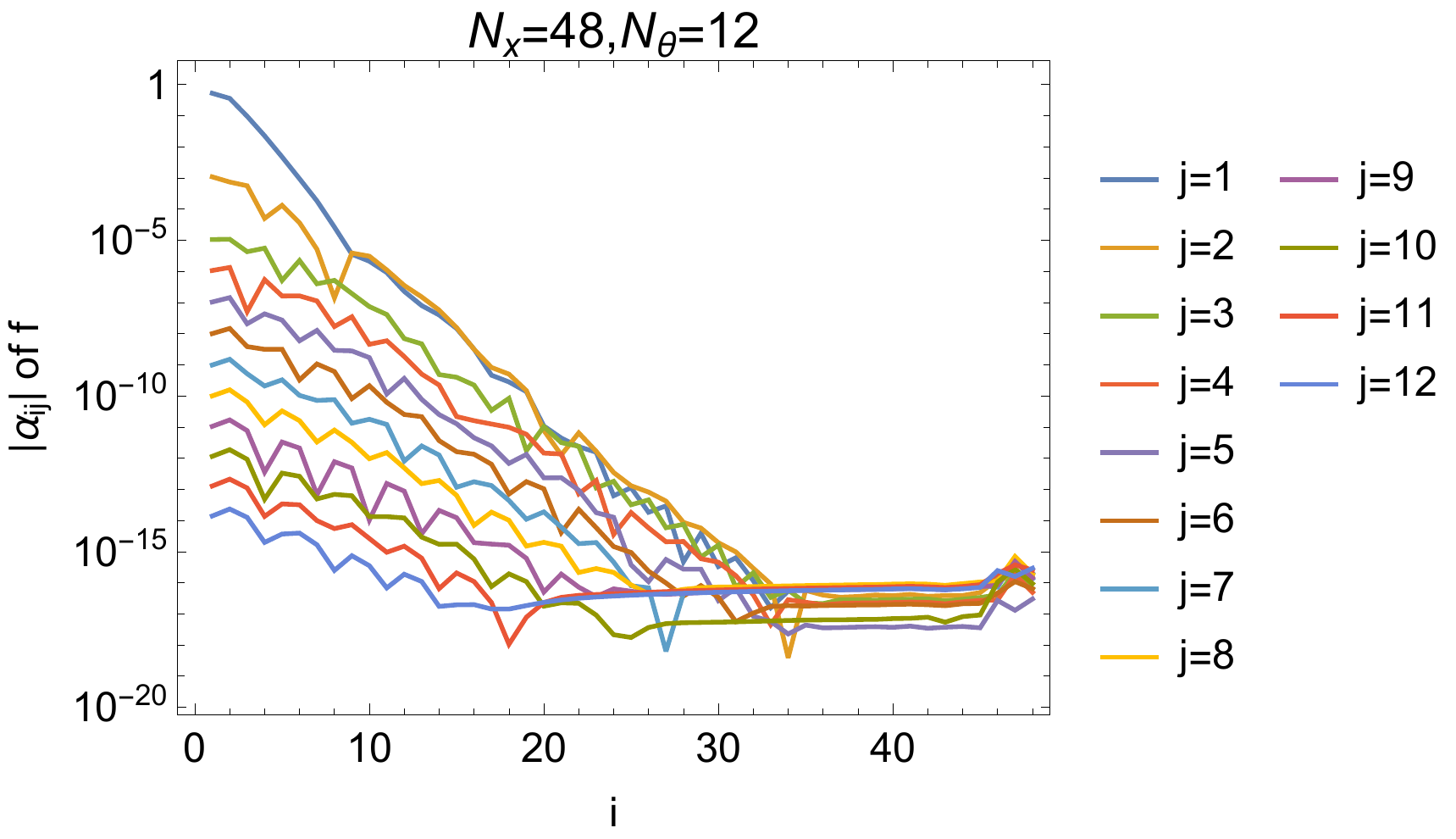}
\caption{The absolute values of the spectral coefficients $\alpha_{ij}$ of the metric function $f(x,\theta)$ with different $N_x$ and $N_\theta$.}\label{fig11}
\end{figure}
At this stage, it would be better to summarize our numerical approach briefly.  To solve the field equations, some preliminary work must be done. First, we
employ the metric ansatz of Eq.(\ref{metric}) which includes five unknown functions, $f$, $g$, $h$, $W$ and $\phi$. Plugging this metric ansatz into the field equations \eqref{eom} and \eqref{KG} leads to a
set of nonlinear coupled PDEs that depend on the functions and their first and second
derivatives $(F, \partial_rF; \partial_r^2F; \partial_\theta F, \partial_\theta^2 F; \partial_{r\theta} F)$.

The set of field equations is then expressed
in terms of the compactified coordinate $x$ defined in Eq.(\ref{rx}) and put in residual form
($R(x,\theta,\partial F)=0$). We do the same thing  for the appropriate boundary conditions. This part of the process could be usually done when  resorting to a computer algebra
system by \textit{Mathematica}. The residuals (and appropriate
Jacobian) are then exported to a coding file  to solve the problem  by using the
developed numerical infrastructure.

Each function is expanded in a spectral series given
by Eq.(\ref{spexpansion}) and the input parameters are then specified (depending on the chosen
boundary conditions for the function $W$). To solve the field equations  successfully, a
good initial guess must be provided to our Newton solver. For this purpose, we interpolate the
functions for  the known Kerr solution using Eq. (\ref{alpha}) and  obtain appropriate spectral
coefficients are provided as a good initial guess.  If new fields are present, we take advantage of perturbative solutions typically and
interpolate them as a guess. Convergence is achieved once the norm difference between spectral coefficients of two successive iterations is less than a certain prescribed
tolerance.
To speed up the solver, the values of our basis functions and their first and second
derivatives are calculated at all grid points and stored, so that no repeated
evaluations are performed. Another optimization
is to store the values on the grid of the trigonometric functions that  appear
in the residuals, $\sin\theta$ and $\cos\theta$.

As an example, the spectral coefficients $\alpha_{ij}$ of the metric function $f$ of a black hole are plotted in Fig. \ref{fig11} with different $N_x$ and $N_\theta$, where the parameters of the black hole are $r_H=0.01$ and $j=0.2$. As can be seen in Fig. \ref{fig11}, the absolute values of the coefficients decrease exponentially as $i$ or $j$ increases, which indicates the convergence of the numerical scheme. Thus, the value of $f$ depends mainly on the first few terms of the coefficients $\alpha_{ij}$. Even if the values of $(N_x,N_\theta)$ are increased from $(40,8)$ to $(44,10)$ and $(48,12)$, the values of the original coefficients do not change and the added coefficients are neglectable. Considering the numerical accuracy and computational resources, we mainly use the setting of $N_x = 40$ and $N_\theta=8$ in the computations in the present paper.

\end{document}